\documentclass[showpacs,amsmath,amssymb,twocolumn]{revtex4}
\usepackage{graphicx}
\usepackage{dcolumn}
\usepackage{bm}

\begin{document}

\title{\bf\large{BEC-BCS Crossover in the Nambu--Jona-Lasinio Model of QCD }}
\author{\normalsize{Gaofeng Sun\footnote{Email address: sgf02@mails.tsinghua.edu.cn},
Lianyi He\footnote{Email address: hely04@mails.tsinghua.edu.cn}
and Pengfei Zhuang\footnote{Email address:
        zhuangpf@mail.tsinghua.edu.cn}}}
\affiliation{Physics Department, Tsinghua University, Beijing
100084, China}

\begin{abstract}
The BEC-BCS crossover in QCD at finite baryon and isospin chemical
potentials is investigated in the Nambu--Jona-Lasinio model. The
diquark condensation in two color QCD and the pion condensation in
real QCD would undergo a BEC-BCS crossover when the corresponding
chemical potential increases. We determined the crossover chemical
potential as well as the BEC and BCS regions. The crossover is not
triggered by increasing the strength of attractive interaction
among quarks but driven by changing the charge density. The chiral
symmetry restoration at finite temperature and density plays an
important role in the BEC-BCS crossover. For real QCD, strong
couplings in diquark and vector meson channels can induce a
diquark BEC-BCS crossover in color superconductor, and in the BEC
region the chromomagnetic instability is fully cured and the
ground state is a uniform phase.
\end{abstract}


\maketitle

\section {Introduction}
\label{s1}
There exists a rich phase structure of Quantum Chromodynamics
(QCD), for instance, the deconfinement process from hadron gas to
quark-gluon plasma and the transition from chiral symmetry
breaking to the symmetry restoration\cite{hwa} at high temperature
and/or baryon density, the color superconductivity at low
temperature but high baryon density\cite{alford,rapp}, and the
pion superfluidity at low temperature but high isospin
density\cite{son}. The physical motivation to study the QCD phase
diagram is closely related to the investigation of the early
universe, compact stars and relativistic heavy ion collisions.

While the perturbation theory of QCD can well describe the
properties of the new phases at high temperature and/or high
density, the study on the phase transitions themselves at moderate
temperature and density depends on lattice QCD calculation and
effective models with QCD symmetries. While there is not yet
precise lattice result for real QCD at finite baryon density due
to the fermion sign problem\cite{karsch}, it is in principle no
problem to do lattice simulation for two color QCD at finite
baryon density\cite{2QCD1,2QCD2,2QCD3,2QCD4} and real QCD at
finite isospin density\cite{ISO1,ISO2,ISO3}. The QCD phase
transitions at finite baryon and/or isospin chemical potential are
also investigated in many low energy effective models, such as
chiral perturbation
theory\cite{CBT1,CBT2,CBT3,CBT4,CBT5,CBT6,CBT7,loewe}, linear
sigma model\cite{LIN1,LIN2,LIN3}, Nambu--Jona-Lasinio
model\cite{NJL1,NJL2,NJL3,NJL4,NJL5,NJL6,NJL7,NJL8}, random matrix
model\cite{MAT1,MAT2}, ladder QCD\cite{LADD}, strong coupling
lattice QCD\cite{LAT}, and global color model\cite{GCM}.

It is generally expected that there would exist a crossover from
Bose-Einstein condensation(BEC) to Bardeen-Cooper-Shriffer
condensation(BCS) for diquarks at finite baryon density and pions
at finite isospin density. In the chemical potential region above
but close to the critical value $\mu_c$ for diquark condensate in
two color QCD or pion condensate in real QCD, since the
deconfinement does not yet happen, the system should be in the BEC
state of diquarks or pions. On the other hand, at a sufficiently
high chemical potential $\mu>>\mu_c$, the ground state of the
system becomes a BCS superfluid where quark-quark or
quark-antiquark cooper pairs are condensed. Therefore, there
should be a crossover from BEC state to BCS state when the
chemical potential or density increases. The BEC-BCS crossover,
which is a hot topic in and beyond condensed matter and ultracold
fermion gas\cite{BEC1,BEC2,BEC3,BEC4,BEC5,BEC6,babaev,qqbar}, was
recently extended to relativistic fermion
superfluid\cite{RBEC1,RBEC2,RBEC3}. In these studies, the BEC-BCS
crossover is induced by increasing the attractive coupling among
fermions. However, this is not the only way to trigger BEC-BCS
crossover. In this paper, we will study the BEC-BCS crossover
induced by density effect and chiral symmetry restoration. This
density induced BEC-BCS crossover is like the one found in nuclear
matter\cite{NM} and non-dilute fermi gas\cite{NDFG}.

The features of BEC-BCS crossover have been widely discussed in
condensed matter and ultracold fermion
gas\cite{BEC1,BEC2,BEC3,BEC4,BEC5,BEC6}. We list here only the
essential characteristics:
\\ 1)There exist two important temperatures, one is the phase transition
temperature $T_c$ for the superfluid at which the order parameter
vanishes, and the other is the molecule dissociation temperature
$T^*$. In the BCS limit, there exists no stable molecule, and
$T^*$ approaches to $T_c$. In the BEC limit, all fermions form
stable difermion molecules, and $T^*$ becomes much larger than
$T_c$.
\\ 2)The chemical potential is equal to the Fermi energy of
non-interacting fermion gas in the BCS limit but becomes negative
in the BEC region. In the BEC limit, the absolute value of the
chemical potential tends to be half of the molecular binding
energy. This is true both near the critical temperature and in the
superfluid ground state.
\\ 3)The fermion momentum distribution changes significantly when we go from the BCS to BEC. In
the BCS limit, the distribution near the Fermi surface is still
very sharp and similar to the one in non-interacting fermion gas.
However, it becomes very smooth in the whole momentum space in the
BEC limit.

To be specific, we express the typical dispersion of fermion
excitations in a superfluid as
\begin{equation}
E_{\bf p}=\sqrt{(\xi_{\bf p}-\mu)^2+\Delta^2},
\end{equation}
where $\Delta$ is the superfluid order parameter. In
non-relativistic homogeneous system, we have $\xi_{\bf p}={\bf
p}^2/(2m)$. In the BCS case, we have $\mu>0$, the minimum of the
dispersion is located at nonzero momentum $|{\bf p}|=\sqrt{2m\mu}$
and the excitation gap is $\Delta$. However, when $\mu$ becomes
negative in the BEC region, the excitation gap becomes
$\sqrt{\mu^2+\Delta^2}$ rather than $\Delta$ itself, and the
minimum of the dispersion is located at $|{\bf p}|=0$. This can be
regarded as the definition of a BEC-BCS crossover at zero
temperature. How can we extend this definition to relativistic
systems? In relativistic case, we have $\xi_{\bf p}=\sqrt{{\bf
p}^2+m^2}$. In comparison with the non-relativistic result, we
should define a new chemical potential $\mu_N=\mu-m$. For $\mu>m$,
we have $\mu_N>0$, the minimum of the dispersion is located at
nonzero momentum $|{\bf p}|=\sqrt{\mu^2-m^2}$ and the excitation
gap is $\Delta$. On the other hand, for $\mu<m$, $\mu_N$ becomes
negative, the excitation gap becomes $\sqrt{\mu_N^2+\Delta^2}$ and
the minimum of the dispersion is located at $|{\bf p}|=0$. This
analysis indicates that the BEC-BCS crossover in relativistic
fermion gas is controlled by $\mu_N=\mu-m$ rather the chemical
potential $\mu$ itself. An interesting phenomenon then arises: The
BEC-BCS crossover would happen when the fermion mass $m$ varies in
the process of chiral symmetry restoration at finite temperature
and density.

The effective models at the hadron level can only describe the BEC
state of hadrons and diquarks\cite{DBEC1,DBEC2}, they can not
describe the possible BEC-BCS crossover when the chemical
potential increases. One of the effective models that enables us
to describe both quark and meson properties is the
Nambu--Jona-Lasinio(NJL) model\cite{nambu} applied to
quarks\cite{vogl,sandy,volkov,hatsuda,buballa}. Even though there
is no confinement in the NJL model, the chiral phase transition
line\cite{vogl,sandy,volkov,hatsuda,buballa,hufner,zhuang} in the
temperature and baryon chemical potential ($T-\mu_B$) plane
calculated in the model is very close to the one obtained from
lattice QCD. It is natural to extend the NJL model to studing
diquark condensation at finite baryon chemical potential and pion
condensation at finite isospin chemical potential. While there is
no reliable lattice result for real QCD with diquark condensation
at finite baryon chemical potential, the calculation of diquark
condensation in the NJL model\cite{NJL5} agrees well with the
lattice simulation of two color QCD\cite{2QCD1,2QCD2}. Also, the
NJL model calculation of pion superfluidity\cite{NJL7} agrees well
with the lattice simulation of real QCD at finite isospin chemical
potential\cite{ISO1,ISO2,ISO3}. It is natural to ask: How can such
a model with quarks as elementary blocks describe the BEC-BCS
crossover of diquarks and pions? In non-relativistic fermion gas,
the microscopic model with four-fermion interaction can describe
well the BEC-BCS crossover at least at zero
temperature\cite{BEC2,BEC5}. Motivated by this fact, we believe
that the NJL model should describe well the BEC-BCS crossover in
QCD at finite density.

The paper is organized as follows. In section \ref{s2} we
investigate the diquark BEC-BCS crossover at finite baryon
chemical potential in the NJL model of two color QCD. The possible
diquark BEC-BCS crossover in real QCD is investigated in the NJL
model in section \ref{s3}. In section \ref{s4} we study the
BEC-BCS crossover of pion condensation at finite isospin chemical
potential. We summarize in section \ref{s5}.

\section {Diquark BEC-BCS Crossover in Two Color NJL Model}
\label{s2}
Let us first consider the diquark condensation in two color QCD.
The advantage to take two color QCD is that the diquarks in this
case are colorless baryons and the diquark condensation breaks the
baryon symmetry U$_{B}(1)$ rather than the color symmetry
SU$_C(2)$. The confinement in two color QCD is less important than
in three color QCD. We start with the two color and two flavor NJL
model
\begin{eqnarray}
\label{njl}
{\cal L} &=&
\bar{\psi}\left(i\gamma^{\mu}\partial_{\mu}-m_0\right)\psi
+G_s\left[\left(\bar{\psi}\psi\right)^2+\left(\bar{\psi}i\gamma_5{\bf \tau}\psi\right)^2\right]\nonumber\\
&&+G_d(\bar{\psi}i\gamma_5\tau_2t_2C\bar{\psi}^{\text{T}})(\psi^{\text{T}}Ci\gamma_5\tau_2t_2\psi),
\end{eqnarray}
where $m_0$ is the current quark mass, $\tau_i$ and $t_i$ are the
Pauli matrices in flavor and color spaces, and the two coupling
constants $G_s$ and $G_d$ are connected by a Fierz transformation
in color space\cite{NJL5}, $G_s=G_d=G$. In the chiral limit, the
Lagrangian (\ref{njl}) with flavor number $N_f=2$ has an enlarged
flavor symmetry SU(2$N_f$), which is called Pauli-Guersey
symmetry\cite{NJL5} and connects quarks and anti-quarks. As a
consequence, diquarks are color singlet baryons and diquarks and
pions become degenerate.

The key quantity describing a thermodynamic system is the
partition function $Z$ which can be expressed as
\begin{equation}
Z=\int[d\bar{\psi}][d\psi]e^{\int_0^\beta d\tau\int d^3{\bf
x}\left({\cal L}+\frac{\mu_B}{2}\bar{\psi}\gamma_0\psi\right)}
\end{equation}
in the imaginary time formulism of finite temperature field
theory, where the baryon chemical potential $\mu_B$ is introduced
explicitly, and $\beta$ is the inverse temperature, $\beta=1/T$.
Using the Hubbard-Stratonovich transformation, we introduce the
auxiliary meson fields $\sigma=-2G\bar{\psi}\psi,\
\pi=-2G\bar{\psi}i\gamma_5{\bf \tau}\psi$ and diquark field
$\phi=-2G\psi^{\text{T}}Ci\gamma_5\tau_2t_2\psi$, and the
partition function can be written as
\begin{equation}
Z=\int[d\bar{\Psi}][d\Psi][d\sigma][d{\bf
\pi}][d\phi][d\phi^*]e^{\int_0^\beta d\tau\int d^3{\bf x}{\cal
L}_{\text{eff}}}
\end{equation}
with the effective Lagrangian
\begin{equation}
{\cal L}_{\text{eff}}=\frac{1}{2}\bar{\Psi}{\cal K}[\sigma,{\bf
\pi},\phi]\Psi-\frac{\sigma^2+{\bf \pi}^2+|\phi|^2}{4G},
\end{equation}
where we have introduced the Nambu-Gorkov spinors
\begin{equation}
\Psi = \left(\begin{array}{cc} \psi\\
C\bar{\psi}^{\text{T}}\end{array}\right),\ \ \ \bar{\Psi} =
\left(\begin{array}{cc} \bar{\psi} &
\psi^{\text{T}}C\end{array}\right),
\end{equation}
and the kernel ${\cal K}$ is defined as
\begin{equation}
{\cal K}[\sigma,{\bf \pi},\phi]=\left(\begin{array}{cc} {\cal M}_+&i\gamma_5\phi\tau_2t_2\\
i\gamma_5\phi^*\tau_2t_2 & {\cal M}_-\end{array}\right)
\end{equation}
with ${\cal
M}_\pm=i\gamma^\mu\partial_\mu-m_0\pm\mu_B\gamma_0/2-(\sigma\pm
i\gamma_5{\bf \tau}\cdot{\bf \pi})$. Integrating out the quark
degrees of freedom, we obtain
\begin{equation}
Z=\int[d\sigma][d{\bf
\pi}][d\phi][d\phi^*]e^{-S_{\text{eff}}[\sigma,{\bf \pi},\phi]}
\end{equation}
with the bosonic effective action
\begin{eqnarray}
\label{eff} S_{\text{eff}}[\sigma,{\bf \pi},\phi]&=&\int_0^\beta
d\tau\int
d^3{\bf x}\frac{\sigma^2+{\bf \pi}^2+|\phi|^2}{4G}\nonumber\\
&&-\frac{1}{2}\text{Tr}\ln{\cal K}[\sigma,{\bf \pi},\phi].
\end{eqnarray}

\subsection {Diquark Fluctuation at $T>T_c$}
The diquarks would be condensed when the baryon chemical potential
$\mu_B$ is larger than their mass $m_d=m_\pi$. In this subsection
we focus on the region above the critical temperature $T_c$ where
the diquark condensate vanishes.

After a field shift $\sigma\rightarrow
\langle\sigma\rangle+\sigma$ with
$\langle\sigma\rangle=-2G\langle\bar{\psi}\psi\rangle$, the
effective action $S^{(0)}_{\text{eff}}$ at zeroth order in meson
and diquark fields gives the mean field thermodynamic potential
$\Omega$,
\begin{eqnarray}
\Omega=\frac{1}{\beta
V}S_{\text{eff}}^{(0)}=\frac{(m-m_0)^2}{4G}+\frac{1}{2\beta
V}\ln\det{\cal S}^{-1},
\end{eqnarray}
where $m=m_0+\langle\sigma\rangle$ is the effective quark mass,
and ${\cal S}$ is the quark propagator at mean field level
\begin{equation}
{\cal S}= \left(\gamma^\mu
\partial_\mu-m+{\mu_B\over2}\gamma_0\sigma_3\right)^{-1}
\end{equation}
with $\sigma_i$ being the Pauli matrices in the Nambu-Gorkov
space. The quadratic term of the effective action reads
\begin{eqnarray}
S_{\text{eff}}^{(2)}[\sigma,\pi,\phi]&=&\int_0^\beta d\tau\int
d^3{\bf x}\frac{\sigma^2+{\bf \pi}^2+|\phi|^2}{4G}\nonumber\\
&&+\frac{1}{4}\text{Tr}\left\{{\cal S}\Sigma[\sigma,{\bf
\pi},\phi]{\cal S}\Sigma[\sigma,{\bf \pi},\phi]\right\},
\end{eqnarray}
where $\Sigma$ is defined as
\begin{eqnarray}
\Sigma[\sigma,{\bf \pi},\phi]=\left(\begin{array}{cc} \sigma+i\gamma_5{\bf \tau}\cdot{\bf \pi}
&i\gamma_5\phi\tau_2t_2\\
i\gamma_5\phi^*\tau_2t_2 & \sigma-i\gamma_5{\bf \tau}\cdot{\bf
\pi}\end{array}\right).
\end{eqnarray}
In momentum space, the second order effective action can be
expressed as
\begin{eqnarray}
S_{\text{eff}}^{(2)}[\sigma,{\bf
\pi},\phi]&=&\frac{1}{2}\sum_k\Bigg[\left[\frac{1}{2G}-\Pi_\sigma(k)\right]
\sigma(-k)\sigma(k)\nonumber\\
&&+\left[\frac{1}{2G}-\Pi_\pi(k)\right]{\bf \pi}(-k)\cdot{\bf \pi}(k)\nonumber\\
&&+\left[\frac{1}{2G}-\Pi_d(k)\right]\phi^*(-k)\phi(k)\nonumber\\
&&+\left[\frac{1}{2G}-\Pi_{\bar
d}(k)\right]\phi(-k)\phi^*(k)\Bigg],
\end{eqnarray}
where $k=(k_0,{\bf k})$ with $k_0=i\omega_n=2in\pi T
(n=0,\pm1,\pm2,\cdots)$ is the meson or diquark four momentum, and
$\sum_k=iT\sum_n\int d^3{\bf k}/(2\pi)^3$ indicates integration
over the three momentum ${\bf k}$ and summation over the frequency
$\omega_n$. The polarization functions for the mesons, diquark and
anti-diquark can be evaluated as
\begin{eqnarray}
&&\Pi_\sigma(k)=2\Pi_1(k;\mu_B),\ \
\Pi_\pi(k)=2\Pi_2(k;\mu_B),\nonumber\\
&&\Pi_d(k)=2\Pi_3(k;\mu_B),\ \ \Pi_{\bar d}(k)=2\Pi_4(k;\mu_B)
\end{eqnarray}
with the functions $\Pi_{1,2,3,4}(k;\mu)$ listed in Appendix A.

Without loss of generality, we consider the case $\mu_B>0$. The
transition temperature $T_c$ for the diquark condensation is
determined by the well-known Thouless criterion
\begin{equation}
\label{tc}
1-2G\Pi_d(k=0)\Big|_{T=T_c}=0
\end{equation}
together with the gap equation for the effective quark mass
derived from the first order derivative of the thermodynamic
potential,
\begin{equation}
\frac{m-m_0}{16Gm}=\int{d^3{\bf p}\over (2\pi)^3}{1-f\left(E_{\bf
p}^+\right)-f\left(E_{\bf p}^-\right)\over E_{\bf p}}
\Bigg|_{T=T_c}
\end{equation}
with the particle energies $E_{\bf p}=\sqrt{{\bf p}^2+m^2}$ and
$E_{\bf p}^\pm=E_{\bf p}\pm\mu_B/2$.

The NJL model is non-renormalizable, a proper regularization is
needed to avoid ultraviolet divergence. For instance, one can add
a form factor in the above momentum integrations. Since our goal
is to study the BEC-BCS crossover which happens far away from the
asymptotic region, we employ, for the sake of simplicity, a hard
three momentum cutoff $\Lambda$ to regularize the above equations.
In the following numerical calculations, we take the current quark
mass $m_0=5$ MeV, the coupling constant $G=1.5\times4.93$
GeV$^{-2}$ and the cutoff $\Lambda=653$ MeV to fit the pion mass
$m_\pi=134$ MeV, the pion decay constant $f_\pi=93$ MeV and the
constituent quark mass $m=300$ MeV in the vacuum.

\begin{figure}[!htb]
\begin{center}
\includegraphics[width=6cm]{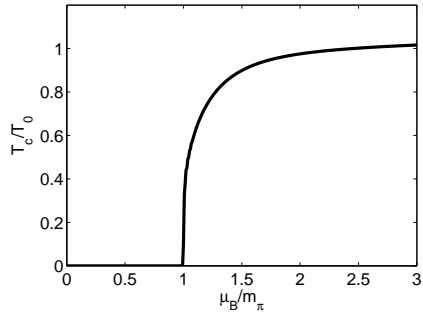}
\caption{The superfluid phase transition temperature $T_c$ scaled
by $T_0$ as a function of baryon chemical potential $\mu_B$ scaled
by $m_\pi$. \label{fig1}}
\end{center}
\end{figure}

The critical temperature as a function of baryon chemical
potential is shown in Fig.\ref{fig1}. The diquark condensed phase
starts at $\mu_B=m_\pi$ and the critical temperature increases
with $\mu_B$ until a saturation value which is about the critical
temperature for chiral symmetry restoration at $\mu_B=0$,
$T_0=185$ MeV. At moderate baryon chemical potential, the critical
temperature can be well described by\cite{splittorff}
\begin{equation}
T_c=T_0\sqrt{1-\left({m_\pi\over \mu_B}\right)^4}.
\end{equation}

\begin{figure}[!htb]
\begin{center}
\includegraphics[width=6cm]{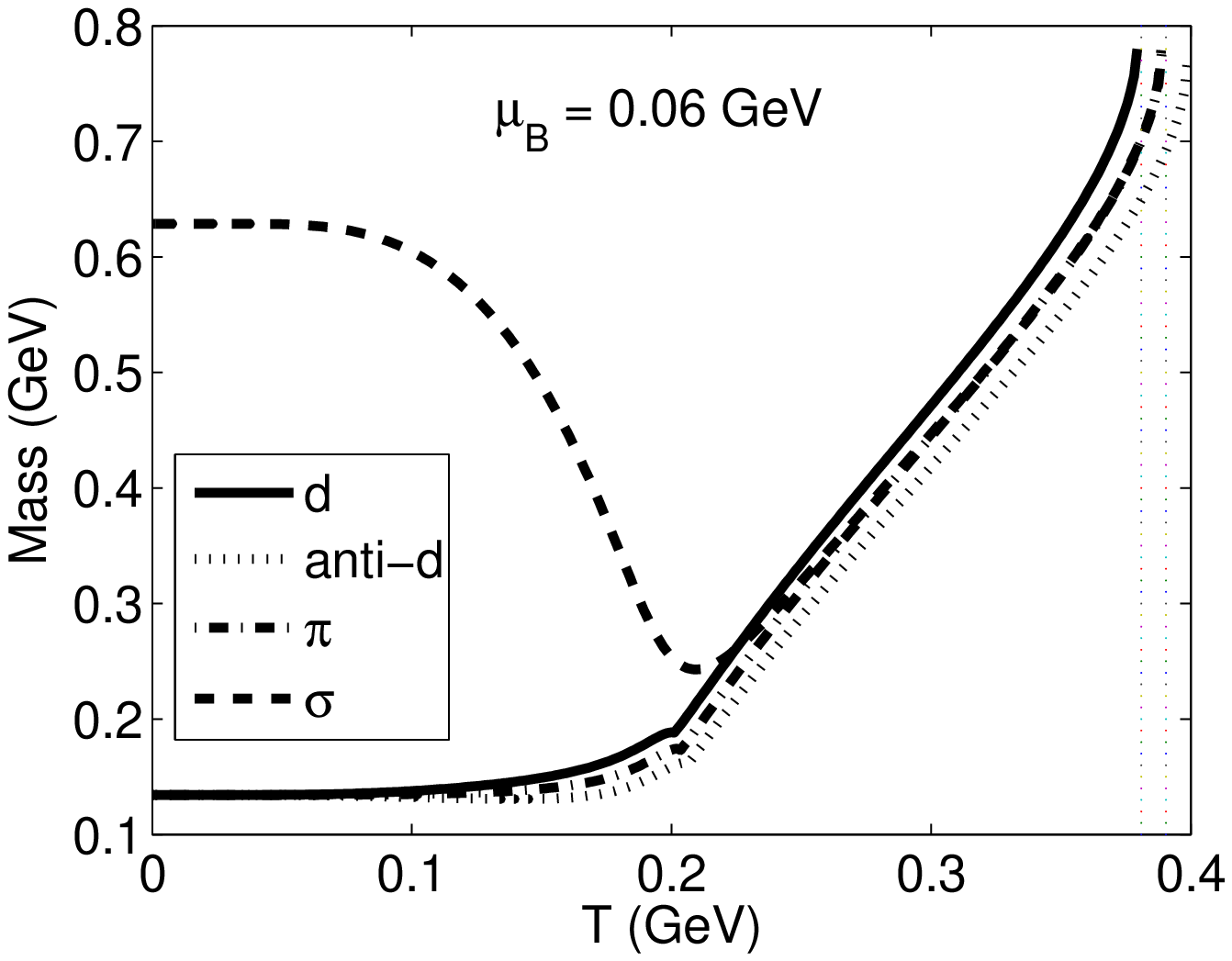}
\includegraphics[width=6cm]{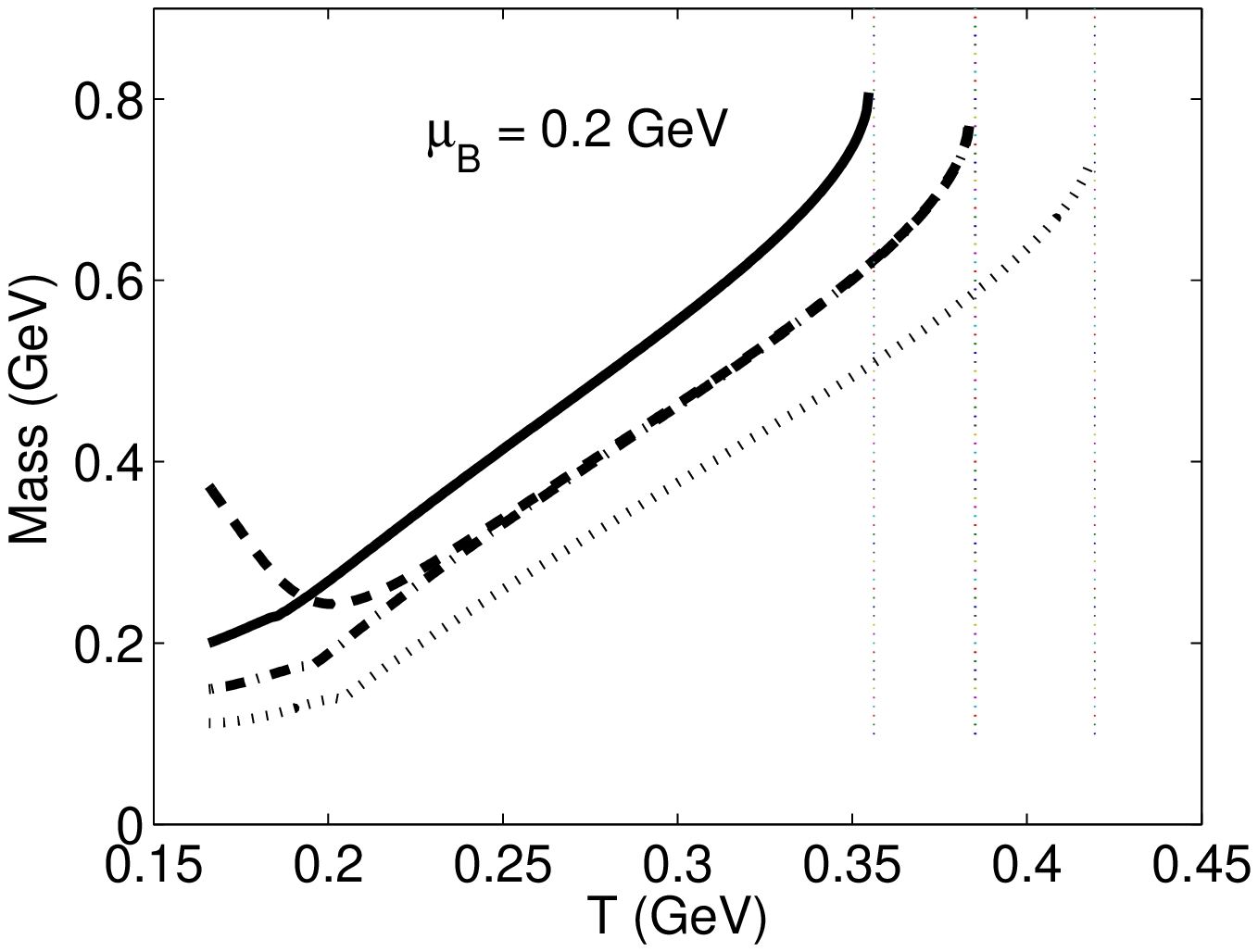}
\caption{The meson and diquark masses as functions of temperature
for $\mu_B=0.06$ GeV and $0.2$ GeV. \label{fig2}}
\end{center}
\end{figure}

\begin{figure}[!htb]
\begin{center}
\includegraphics[width=6cm]{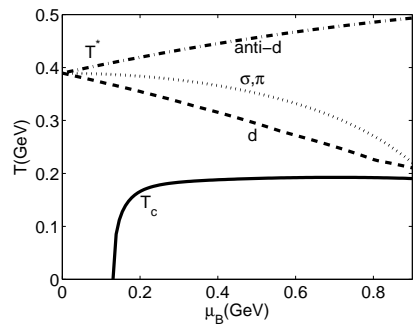}
\caption{The meson and diquark dissociation temperatures as
functions of the baryon chemical potential. \label{fig3}}
\end{center}
\end{figure}

We discuss now the mesons and diquarks above $T_c$. The energy
dispersions of the mesons and diquarks are defined by the poles of
their propagators,
\begin{equation}
\label{pole}
1-2G\Pi_i(\omega({\bf k}),{\bf k})=0,\ \ \
i=\sigma,\pi,d,\bar d.
\end{equation}
While the full dispersion laws can be obtained numerically, we are
interested here only in the meson and diquark masses. For $\sigma$
and ${\bf \pi}$,  they do not carry baryon number, and their
masses are defined as $m_\sigma=\omega_\sigma(0)$ and
$m_{\pi}=\omega_{\pi}(0)$. However, diquarks and anti-diquarks
carry baryon numbers and their masses are not exactly their
dispersions at ${\bf k}=0$. To define a proper diquark or
anti-diquark mass, we should subtract the corresponding baryon
chemical potential from the dispersion, $m_d=\omega_d(0)+\mu_B$
and $m_{\bar d}=\omega_{\bar d}(0)-\mu_B$. For $\mu_B>0$, by
comparing equation (\ref{tc}) for the transition temperature $T_c$
with the above definition of diquark mass, the diquark mass at
$T_c$ is equal to the baryon chemical potential,
\begin{equation}
m_d(T_c)=\mu_B,
\end{equation}
which is consistent with the physical picture of relativistic BEC
in boson field theory. In Fig.\ref{fig2} we show the meson and
diquark mass spectrum above the critical temperature for
$\mu_B<m_\pi$ and $\mu_B>m_\pi$. For $\mu_B<m_\pi$, the meson mass
spectrum is similar to the case at $\mu_B=0$, but the diquark mass
is slightly different from the anti-diquark mass at high
temperature. For $\mu_B>m_\pi$, the mass spectrum starts at $T_c$
where the diquark mass is exactly equal to $\mu_B$. For any
$\mu_B$, we found that the meson or diquark mass becomes divergent
at a limit temperature $T^*$ which indicates the dissociation of
the meson or diquark resonances. For small baryon chemical
potential, this dissociation temperature is about two times the
critical temperature for chiral symmetry restoration, which is
consistent with the results in \cite{he1}. With increasing baryon
chemical potential, while the dissociation temperature for
diquarks and mesons decreases, the one for anti-diquarks
increases. Since the diquarks are condensed at positive $\mu_B$,
the baryon chemical potential dependence of the diquark
dissociation temperature indicates a crossover from diquark BEC to
BCS superfluidity. In Fig.\ref{fig3}, we show the superfluid
transition temperature $T_c$ and the dissociation temperature
$T^*$ for mesons, diquarks and anti-diquarks. Any $T^*$ is much
larger than $T_c$ at small $\mu_B$, but for mesons and diquarks
the two temperatures tend to coincide with each other at
sufficiently large $\mu_B$.

To investigate the diquark excitation in detail,  we consider the
diquark spectral function
\begin{equation}
\rho(\omega,{\bf k})=-2\text{Im}{\text D}_R(\omega,{\bf k}),
\end{equation}
where ${\text D}_R(\omega,{\bf k})\equiv {\cal
D}(\omega+i\eta,{\bf k})$ is the analytical continuation of the
diquark Green function ${\cal D}(i\omega_n,{\bf k})$ defined as
\begin{equation}
{\cal D}(i\omega_n,{\bf k})=\frac{2G}{1-2G\Pi_d(i\omega_n,{\bf
k})}.
\end{equation}
The spectral function $\rho$ can be rewritten as
\begin{eqnarray}
&&\rho(\omega,{\bf k})=\\
&&\frac{-8G^2\text{Im}\Pi_d(\omega+i\eta,{\bf
k})}{\left[1-2G\text{Re}\Pi_d(\omega+i\eta,{\bf
k})\right]^2+\left[2G\text{Im}\Pi_d(\omega+i\eta,{\bf
k})\right]^2}.\nonumber
\end{eqnarray}
At zero momentum ${\bf k}=0$, the real and imaginary parts of the
diquark polarization function can be evaluated as
\begin{eqnarray}
&&\text{Re}\Pi_d(\omega+i\eta,{\bf 0})=\\
&&-8\int{d^3{\bf p}\over (2\pi)^3}\left[\frac{1-2f(E_{\bf
p}^-)}{\omega-2E_{\bf p}+\mu_B}-\frac{1-2f(E_{\bf
p}^+)}{\omega+2E_{\bf p}+\mu_B}\right],\nonumber\\
&&\text{Im}\Pi_{\text{d}}(\omega+i\eta,{\bf 0})=\nonumber\\
&&-8\pi\int{d^3{\bf p}\over (2\pi)^3}\Big[(1-2f(E_{\bf
p}^-))\delta(\omega-2E_{\bf p}+\mu_B)\nonumber\\
&&\ \ \ \ \ \ \ \ \ \ \ \ \ \ \ \ \ \ \ \ -(1-2f(E_{\bf p}^+))\delta(\omega+2E_{\bf p}+\mu_B)\Big]\nonumber\\
&&=\frac{2}{\pi}\left[p_\omega
E_{p_\omega}\left(1-2f(E_{p_\omega}^-)\right)\right]\Theta(\omega+\mu_B-2m),\nonumber
\end{eqnarray}
with
\begin{equation}
p_\omega=\sqrt{\left(\frac{\omega+\mu_B}{2}\right)^2-m^2}.
\end{equation}

\begin{figure}[!htb]
\begin{center}
\includegraphics[width=6cm]{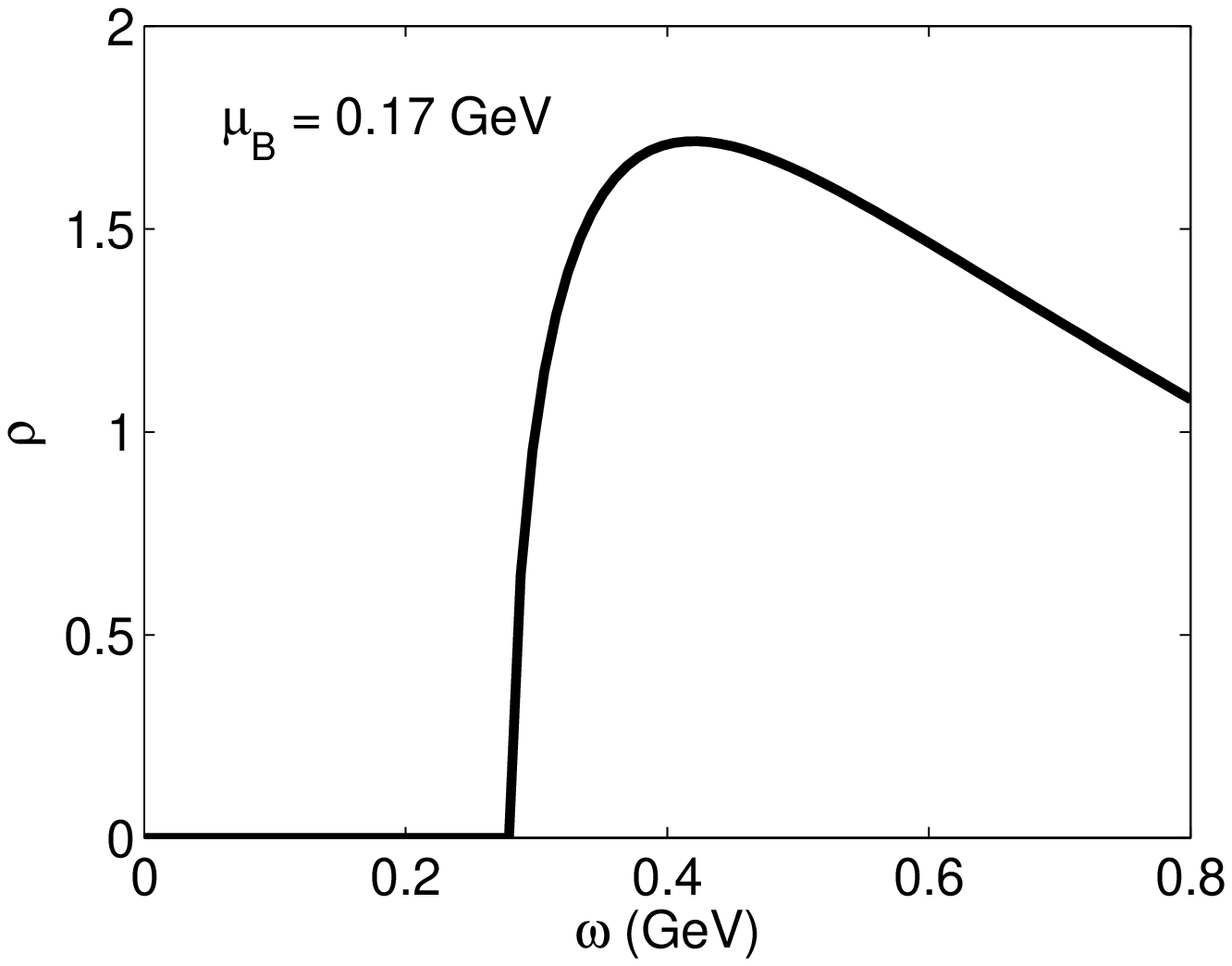}
\includegraphics[width=6cm]{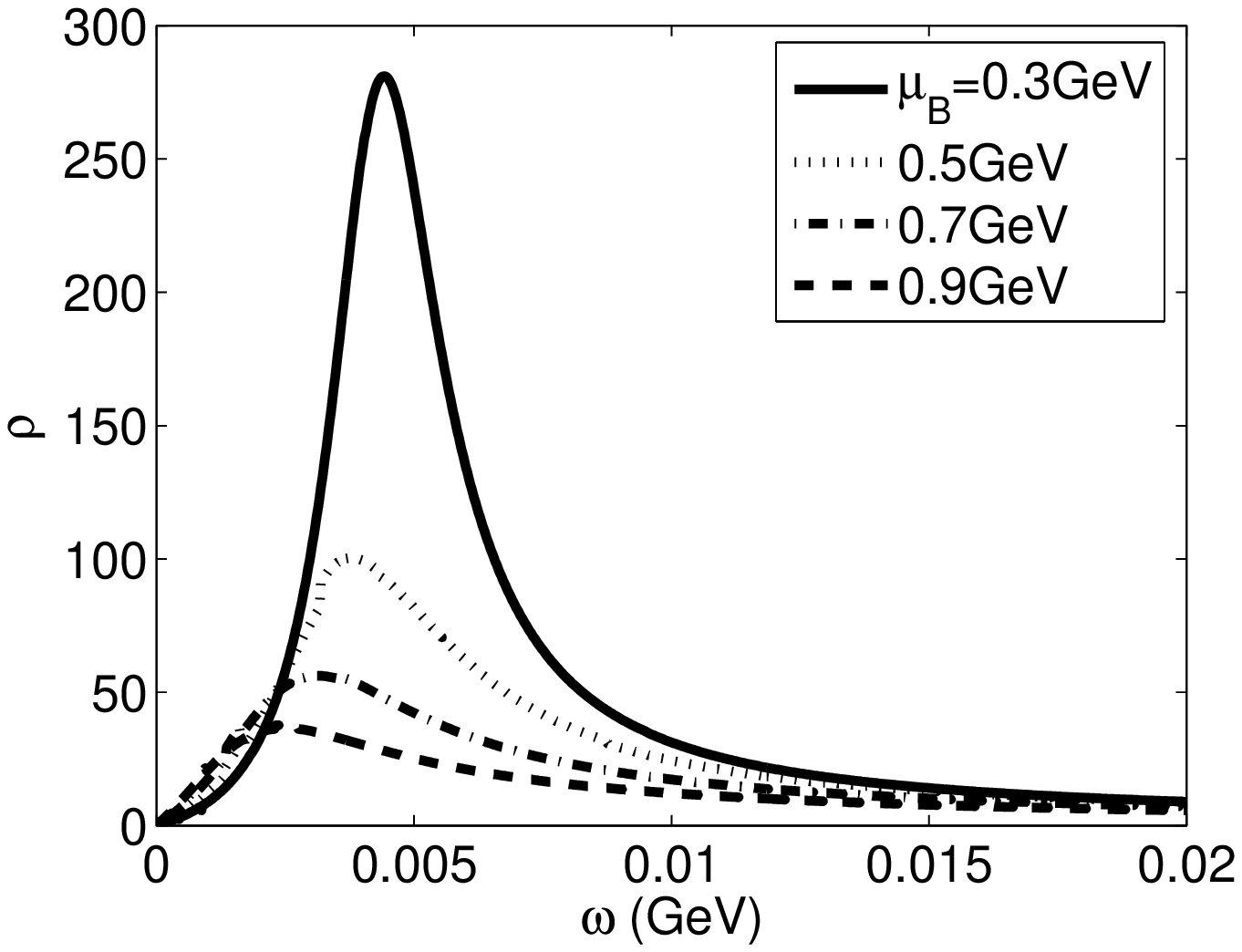}
\caption{The scaled diquark spectral function at zero momentum for
small and large baryon chemical potentials. \label{fig4}}
\end{center}
\end{figure}

For $\mu_B<2m$, a diquark can decay into two quarks only when its
energy satisfies $\omega>2m-\mu_B$, due to the step function
$\Theta$ in the imaginary part. However, for $\mu_B>2m$, a diquark
can decay into two quarks at any energy and hence becomes an
unstable resonance. In Fig.\ref{fig4} we show the diquark spectral
function at zero momentum for several values of baryon chemical
potential near the critical temperature $T_c$. For small baryon
chemical potential $\mu_B<2m$, the diquarks are in stable bound
state at small energy $\omega<2m-\mu_B$ and in unstable resonant
state at large $\omega>2m-\mu_B$. For large baryon chemical
potential $\mu_B>2m$, the diquarks are impossible to stay in bound
state. With increasing baryon chemical potential, the diquarks
become more and more unstable and finally disappear. This
naturally supports the picture of diquark BEC-BCS crossover.
\begin{figure}[!htb]
\begin{center}
\includegraphics[width=6cm]{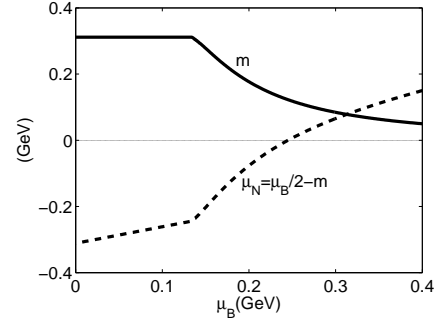}
\caption{The effective quark mass and the corresponding chemical
potential $\mu_N$ as functions of baryon chemical potential at the
critical temperature. \label{fig5}}
\end{center}
\end{figure}

To understand more clearly the crossover from diquark BEC to BCS
superfluidity, we come back to the BEC-BCS crossover in
non-relativistic condensed matter physics. A signal of the
crossover is that in the BEC region the chemical potential becomes
negative and its absolute value tends to be half of the molecule
binding energy. Therefore, the fermions are heavy and hard to be
excited even at finite temperature, and only molecules survive and
condense in the BEC region. In relativistic systems, we should
subtract the fermion mass from the chemical potential in order to
compare with the non-relativistic result. For this purpose, we
define a new chemical potential
\begin{equation}
\mu_N=\frac{\mu_B}{2}-m.
\end{equation}
In the deep BCS limit with $\mu_B\rightarrow\infty$, we have
$m\rightarrow 0$ and $\mu_N\rightarrow\mu_B/2$, but in the deep
BEC limit with $\mu_B\rightarrow m_\pi$, $m$ tends to be its
vacuum value, and $\mu_N$ becomes negative and its absolute value
approaches to half of the diquark binding energy
$\varepsilon_b=2m-m_\pi$. In Fig.\ref{fig5} we show the effective
quark mass $m$ and the chemical potential $\mu_N$ at the critical
temperature. The effective quark mass decreases gradually from its
vacuum value to zero with increasing baryon chemical potential.
Since the lowest quark excitation energy is $\sqrt{{\bf
p}^2+m^2}-\mu_B/2$, the quarks are hard to be excited near the
critical temperature if $m$ is much larger than $\mu_B/2$.
Obviously, at small $\mu_B$, the diquarks are easy to be excited
near $T_c$. This means that even though the elementary blocks of
the NJL model are quarks, the true physical picture at small
baryon chemical potential is the BEC of diquarks. The baryon
chemical potential $\mu_B^0$ corresponding to the crossover can be
defined as $\mu_B=2m$ which is just the point where the diquarks
become unstable resonances. From Fig.\ref{fig5}, $\mu_B^0$ is
about $240$ MeV. We will give an analytical expression for
$\mu_{\text B}^0$ in the next subsection.

\subsection {Diquark Condensation at $T<T_c$}
In this section we focus on the quark and diquark behavior in the
superfluid phase. We start from the effective action (\ref{eff}).
Since diquarks are condensed, we should introduce not only the
chiral condensate $\langle\sigma\rangle$ but also the diquark
condensates,
\begin{equation}
\langle\phi^*\rangle ={\Delta\over \sqrt 2}e^{i\theta}, \ \ \ \ \
\langle\phi\rangle  ={\Delta\over \sqrt 2}e^{-i\theta}.
\end{equation}
A nonzero diquark condensate $\Delta\neq 0$ means spontaneous
breaking of U$_B(1)$ baryon number symmetry in the two color QCD.
The phase factor $\theta$ indicates the direction of the U$_B(1)$
symmetry breaking. For a homogeneous superfluid, we can choose
$\theta=0$ without loss of generality. A gapless Goldstone boson
will appear, which can be identified as the quantum fluctuation in
the phase direction. This Goldstone boson is just the Bogoliubov
phonon in the superfluid.

After a field shift $\sigma\rightarrow
\langle\sigma\rangle+\sigma$ and $\phi\rightarrow \Delta+\phi$,
the effective action at zeroth order in meson and diquark fields
gives the mean field thermodynamic potential
\begin{equation}
\Omega=\frac{1}{\beta
V}S_{\text{eff}}^{(0)}=\frac{(m-m_0)^2+\Delta^2}{4G}+\frac{1}{2\beta
V}\ln\det{\cal S}_\Delta^{-1},
\end{equation}
where ${\cal S}_\Delta$ is the mean field quark propagator in the
superfluid phase. In momentum space it can be evaluated as
\begin{equation}
{\cal S}_\Delta(p)= \left(\gamma^\mu
p_\mu-m+{\mu_B\over2}\gamma_0\sigma_3+i\gamma_5\Delta\tau_2
t_2\sigma_1\right)^{-1}
\end{equation}
with the quark four momentum $p=(i\nu_n,{\bf p})=((2n+1)i\pi
T,{\bf p})\ (n=0,\pm1,\pm2,\cdots)$. In the Nambu-Gorkov space, it
can be explicitly expressed as a matrix
\begin{equation}
\label{smean}
{\cal S}_\Delta(p)= \left(\begin{array}{cc} {\cal S}_{11}(p)&{\cal S}_{12}(p)\\
{\cal S}_{21}(p)&{\cal S}_{22}(p)\end{array}\right)
\end{equation}
with the elements
\begin{eqnarray}
&& {\cal S}_{11} = {\left(i\nu_n+E_{\bf
p}^-\right)\Lambda_+\gamma_0\over
(i\nu_n)^2-(E_\Delta^-)^2}+ {\left(i\nu_n-E_{\bf p}^+\right)\Lambda_-\gamma_0\over (i\nu_n)^2-(E_\Delta^+)^2},\nonumber\\
&& {\cal S}_{22} = {\left(i\nu_n-E_{\bf
p}^-\right)\Lambda_-\gamma_0\over
(i\nu_n)^2-(E_\Delta^-)^2}+ {\left(i\nu_n+E_{\bf p}^+\right)\Lambda_+\gamma_0\over (i\nu_n)^2-(E_\Delta^+)^2},\nonumber\\
&& {\cal S}_{12} = {-i\Delta\tau_2t_2\Lambda_+\gamma_5\over
(i\nu_n)^2-(E_\Delta^-)^2}+ {-i\Delta\tau_2t_2\Lambda_-\gamma_5\over (i\nu_n)^2-(E_\Delta^+)^2},\nonumber\\
&& {\cal S}_{21} = {-i\Delta\tau_2t_2\Lambda_-\gamma_5\over
(i\nu_n)^2-(E_\Delta^-)^2}+
{-i\Delta\tau_2t_2\Lambda_+\gamma_5\over
(i\nu_n)^2-(E_\Delta^+)^2},
\end{eqnarray}
where $\Lambda_\pm$ are the energy projectors
\begin{equation}
\Lambda_{\pm}({\bf p}) = {1\over
2}\left[1\pm{\gamma_0\left(\vec{\gamma}\cdot{\bf p}+m\right)\over
E_{\bf p}}\right],
\end{equation}
and $E_\Delta^\pm = \sqrt{(E_{\bf p}^\pm)^2+\Delta^2}$ are quark
energies in the superfluid phase. The momentum distributions of
quarks and anti-quarks can be calculated from the positive and
negative energy components of the diagonal propagators ${\cal
S}_{11}$ and ${\cal S}_{22}$,
\begin{eqnarray}
n_q({\bf p})=\sum_n{i\nu_n+E_{\bf p}^-\over
(i\nu_n)^2-(E_\Delta^-)^2}e^{i\nu_n\eta}
=\frac{1}{2}\left(1-\frac{E_{\bf p}^-}{E_\Delta^-}\right),\nonumber\\
n_{\bar q}({\bf p})=\sum_n{i\nu_n+E_{\bf p}^+\over
(i\nu_n)^2-(E_\Delta^+)^2}e^{i\nu_n\eta}=\frac{1}{2}\left(1-\frac{E_{\bf
p}^+}{E_\Delta^+}\right).
\end{eqnarray}

\begin{figure}[!htb]
\begin{center}
\includegraphics[width=6cm]{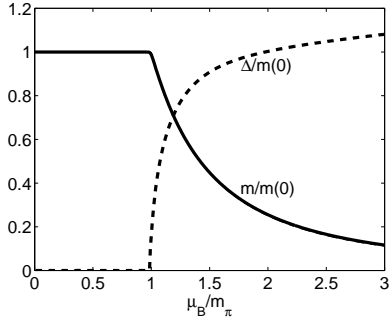}
\caption{The diquark condensate and the effective quark mass
scaled by the quark mass in the vacuum as functions of
$\mu_B/m_\pi$ at zero temperature. \label{fig6}}
\end{center}
\end{figure}

It has been demonstrated in non-relativistic models that the BCS
mean field theory can describe well the BEC-BCS crossover at low
temperature, $T\ll T_c$. Here we will treat the ground state in
the mean field approximation. The gap equations to determine the
effective quark mass $m$ and diquark condensate $\Delta$ can be
obtained by the minimum of the thermodynamic potential,
\begin{equation}
\label{gap} \frac{\partial\Omega}{\partial m}=0,\ \ \
\frac{\partial\Omega}{\partial\Delta}=0.
\end{equation}
With the explicit form of the thermodynamic potential
\begin{equation}
\Omega=\frac{(m-m_0)^2+\Delta^2}{4G}-8\int {d^3 {\bf p}\over
(2\pi)^3}\left[\zeta(E_\Delta^+)+\zeta(E_\Delta^-)\right],
\end{equation}
where $\zeta$ is defined as $\zeta(x)=
x/2+\beta^{-1}\ln(1+e^{-\beta x})$, we obtain the gap equations at
zero temperature
\begin{eqnarray}
\label{gap0} m-m_0 &=& 8Gm\int{d^3{\bf p}\over (2\pi)^3}{1\over
E_{\bf p}} \left({E_{\bf p}^-\over E_\Delta^-}+{E_{\bf p}^+\over
E_\Delta^+}\right) ,\nonumber\\
\Delta &=& 8G\Delta\int{d^3{\bf p}\over (2\pi)^3}\left({1\over
E_\Delta^-}+{1\over E_\Delta^+}\right).
\end{eqnarray}
Numerical solution of the gap equations is shown in
Fig.\ref{fig6}. Our results for the chiral and diquark condensates
agree quite well with the lattice data\cite{NJL5}. For
$\mu_B<m_\pi$ the ground state is the same as the vacuum state and
the baryon density keeps zero. For $\mu_B>m_\pi$ the diquark
condensate and baryon density become nonzero. The critical baryon
chemical potential $\mu_{\text B}^c$ is exactly the diquark mass
$m_d$ in the vacuum, and the phase transition is of second order.
Since the chiral symmetry is spontaneously broken at small
$\mu_{\text B}$ and almost restored at large $\mu_{\text B}$, the
effective quark mass $m$ plays an important role at small
$\mu_{\text B}$ but can be neglected at large $\mu_{\text B}$. In
fact, in the region above but close to the critical value
$\mu_B^c$, the chiral condensate, effective quark mass and the
diquark condensate as functions of $\mu_B$ can be well described
by\cite{NJL7}
\begin{eqnarray}
\label{condensate}
&&\frac{m(\mu_B)}{m(0)}\simeq\frac{\langle\sigma\rangle(\mu_B)}
{\langle\sigma\rangle(0)}=\left(\frac{m_\pi}{\mu_B}\right)^2,
\nonumber\\
&&\frac{\Delta(\mu_B)}{\langle\sigma\rangle(0)}=\sqrt{1-\left(\frac{m_\pi}{\mu_B}\right)^4},
\end{eqnarray}
which are consistent with the results from the chiral effective
theory\cite{CBT2}.

\begin{figure}[!htb]
\begin{center}
\includegraphics[width=6cm]{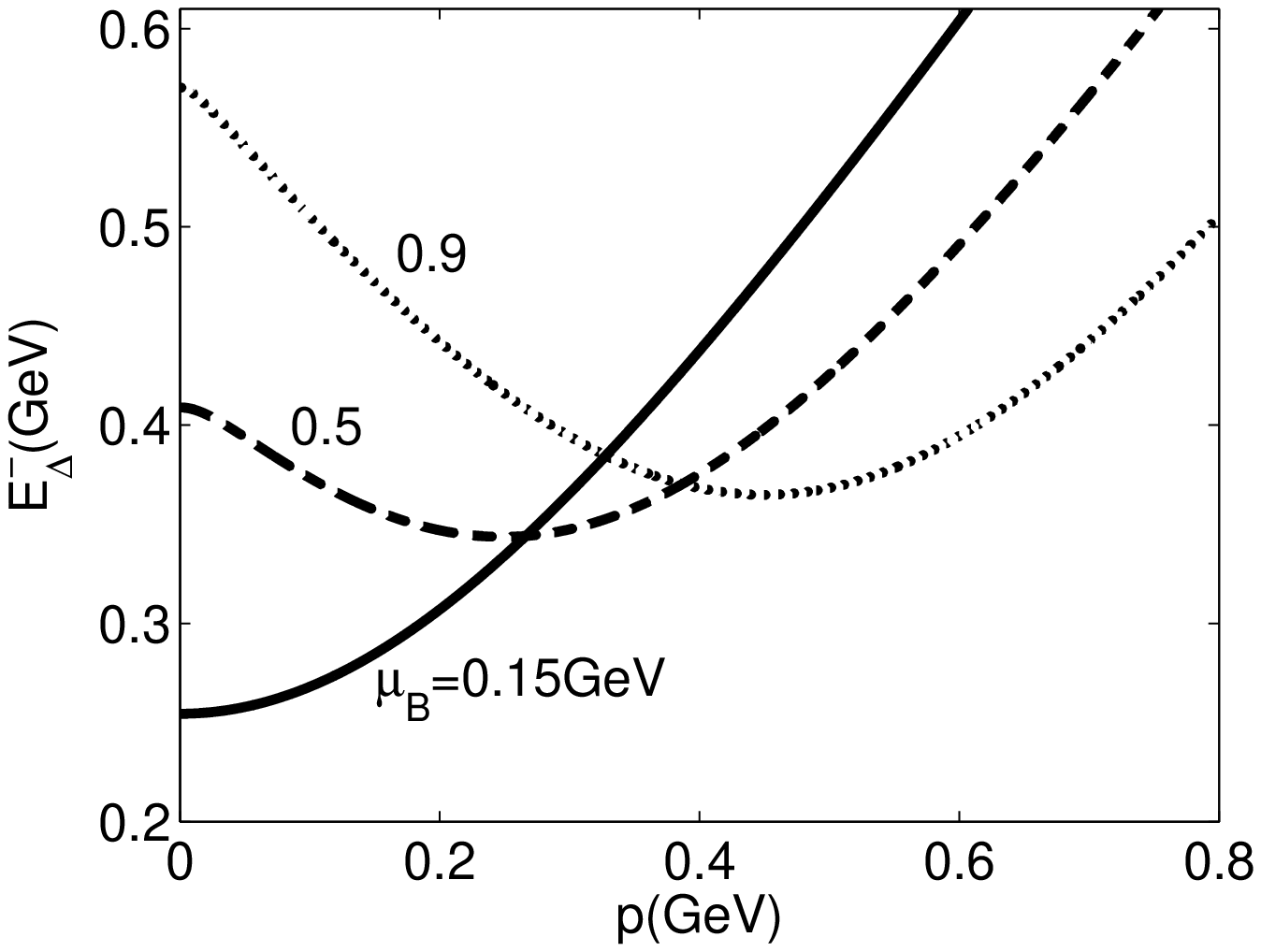}
\includegraphics[width=6cm]{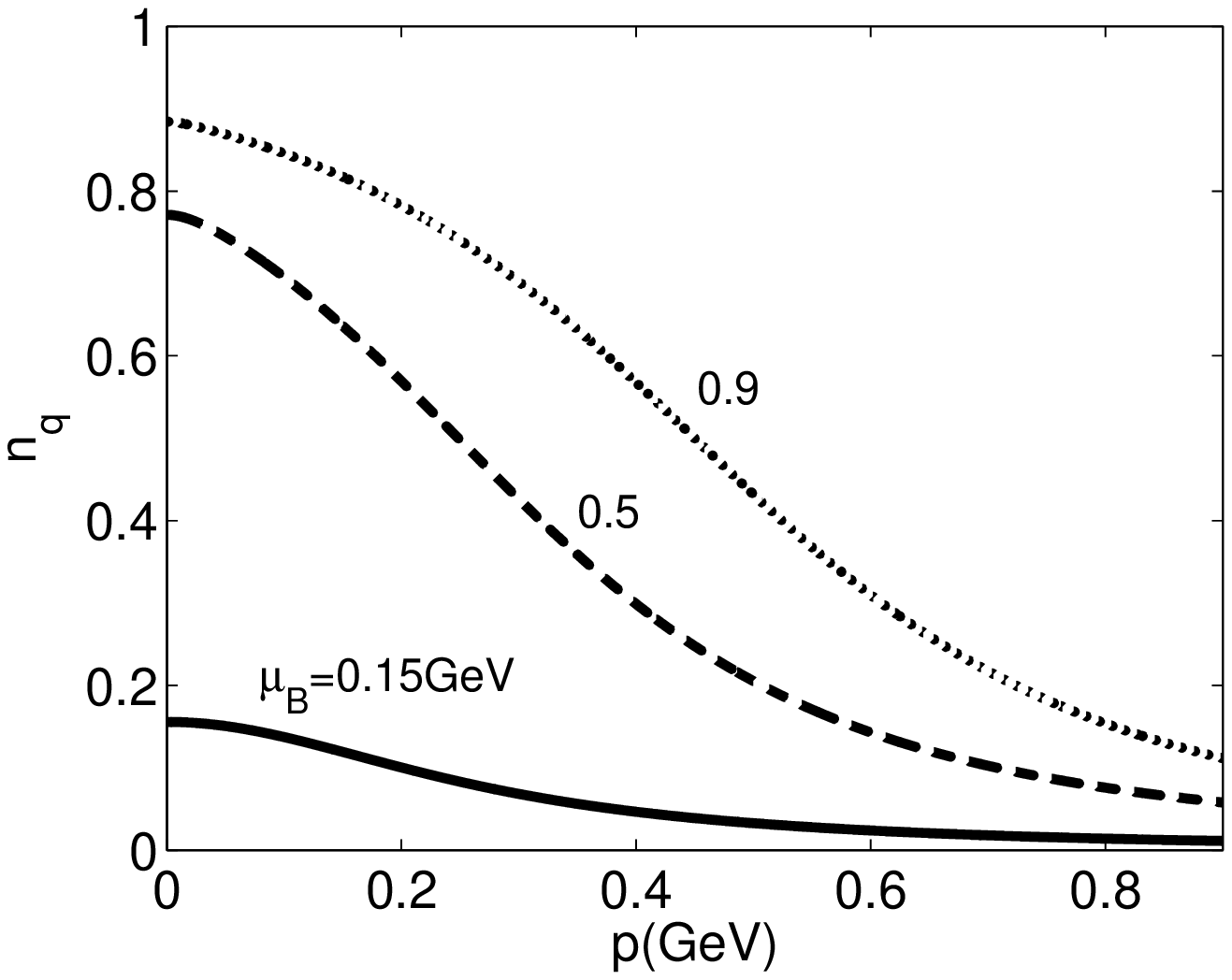}
\caption{The quark energy dispersion and momentum distribution at
several baryon chemical potentials at zero temperature.
\label{fig7}}
\end{center}
\end{figure}

To explain the BEC-BCS crossover, we further consider the
dispersions of quark excitations,
\begin{equation}
E_\Delta^\pm=\sqrt{\left(\sqrt{{\bf
p}^2+m^2}\pm\mu_B/2\right)^2+\Delta^2}.
\end{equation}
In the case of $\mu_B>0$, $E_\Delta^+$ is the anti-particle
excitation. In Fig.\ref{fig7} we show the dispersion $E_\Delta^-$
at $\mu_B=0.15,\ 0.5,\ 0.9$ GeV. Obviously, the fermion
excitations are always gapped. At small $\mu_B$ with
$\mu_B/2<m(\mu_B)$, the minimum of the dispersion is at $|{\bf
p}|=0$ where the energy gap is $\sqrt{\mu_N^2+\Delta^2}$ with the
corresponding non-relativistic chemical potential
$\mu_N=\mu_B/2-m$ introduced in the last subsection. However, at
large $\mu_B$ with $\mu_B/2>m(\mu_B)$, the minimum of the
dispersion is shifted to $|{\bf p}|\simeq\mu_B/2$ where the energy
gap is $\Delta$. Such a phenomenon is a signal of the BEC-BCS
crossover.

The BEC-BCS crossover is also reflected in the momentum dependence
of the quark occupation number. In Fig.\ref{fig7} we show also the
quark momentum distribution $n_q({\bf p})$ at $\mu_B=0.15,\ 0.5,\
0.9$ GeV. At small $\mu_B$ such as $\mu_B=0.15$ GeV the occupation
number is very small and smooth in the whole momentum region,
while at large $\mu_{\text B}$ the occupation number becomes large
near $|{\bf p}|=0$ and drops down with increasing momentum
rapidly. When $\mu_{\text B}$ becomes very large, the occupation
number is indeed of the BCS type.

\begin{figure}[!htb]
\begin{center}
\includegraphics[width=6cm]{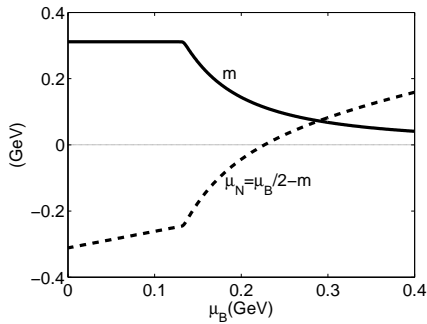}
\caption{The effective quark mass and non-relativistic chemical
potential $\mu_N$ as functions of baryon chemical potential at
zero temperature. \label{fig8}}
\end{center}
\end{figure}

The numerical results of the chemical potential $\mu_N$ is shown
in Fig.\ref{fig8}. The turning point from negative $\mu_N$ to
positive $\mu_N$ is at about $\mu_B^0\simeq 230$ MeV, which is
nearly the same as we obtained at the critical temperature shown
in the last subsection. In fact, we can get an analytical
expression for $\mu_B^0$ using equation (\ref{condensate}). The
BEC region is obtained by requiring $\mu_N<0$, namely
\begin{equation}
\frac{\mu_B}{2}<m(\mu_B)=m(0)\left(\frac{m_\pi}{\mu_B}\right)^2,
\end{equation}
from which we obtain
\begin{equation}
\mu_B^0=\left[2m(0)m_\pi^2\right]^{1/3},
\end{equation}
it depends on the pion mass and effective quark mass in the
vacuum. With the above chosen parameter set, we have $\mu_B^0=230$
MeV. When the effective quark mass in the vacuum varies from $300$
MeV to $500$ MeV, $\mu_B^0$ changes from $230$ MeV to $270$ MeV.
Note that the explicit chiral symmetry breaking induced by nonzero
current quark mass $m_0$ plays an important role in the study of
BEC-BCS crossover. In the chiral limit with $m_0=0$, pions are
massless and there will be no BEC state.

We now turn to the meson and diquark excitations. The second order
effective action which is quadratic in the meson and diquark
fields can be expressed as
\begin{eqnarray}
S_{\text{eff}}^{(2)}[\sigma,{\bf \pi},\phi]&=&\int_0^\beta
d\tau\int
d^3{\bf x}\frac{\sigma^2+{\bf \pi}^2+|\phi|^2}{4G}\\
&&+\frac{1}{4}{\text Tr}\left\{{\cal S}_\Delta\Sigma[\sigma,{\bf
\pi},\phi]{\cal S}_\Delta\Sigma[\sigma,{\bf
\pi},\phi]\right\},\nonumber
\end{eqnarray}
and can be evaluated as
\begin{equation}
S_{\text{eff}}^{(2)}[\sigma,{\bf
\pi},\phi]=\frac{1}{2}\sum_k\left[\frac{\delta_{ij}}{2G}-\Pi_{ij}(k)\right]\varphi_i(-k)\varphi_j(k)
\end{equation}
in momentum space, where $\varphi_i$ stand for the meson and
diquark fields $\sigma, {\bf \pi}, \phi$ and $\phi^*$, and the
polarization functions $\Pi_{ij}$ are defined as
\begin{equation}
\Pi_{ij}(k)=i\int\frac{d^4
p}{(2\pi)^4}\text{Tr}\left[\Gamma_i{\cal
S}_\Delta(p+k)\Gamma_j{\cal S}_\Delta(p)\right]
\end{equation}
with $\Gamma_i$ being the interaction vertex defined in the
Lagrangian density (\ref{njl}).

Due to the off-diagonal elements of the quark propagator
(\ref{smean}), those eigen modes of the system above $T_c$ are, in
general case, no longer the eigen modes of the Hamiltonian in the
superfluid phase. The new eigen modes are linear combinations of
the old eigen modes and their masses are controlled by the
determinant of the meson and diquark polarizations
\begin{equation}
\det\left[\frac{\delta_{ij}}{2G}-\Pi_{ij}(k_0=M,{\bf
k}=0)\right]=0.
\end{equation}
It can be analytically proven that while pions do not mix with the
others and hence are still the eigen modes of the system, there is
indeed a mixing among sigma, diquark and anti-diquark\cite{NJL5},
and this mixing leads to a gapless Goldstone boson. We can show
that $\Pi_{\sigma d}$ and $\Pi_{\sigma \bar d}$ are proportional
to $m\Delta$ and $\Pi_{d \bar d}$ is proportional to $\Delta^2$.
Therefore, the mixing between sigma and diquark or anti-diquark is
very strong in the BEC region where $m$ and $\Delta$ coexist and
are both large, but the mixing can be neglected at large baryon
chemical potential where $m$ approaches to zero.

\subsection {Phase Diagram in $T-\mu_B$ Plane}
\begin{figure}[!htb]
\begin{center}
\includegraphics[width=6cm]{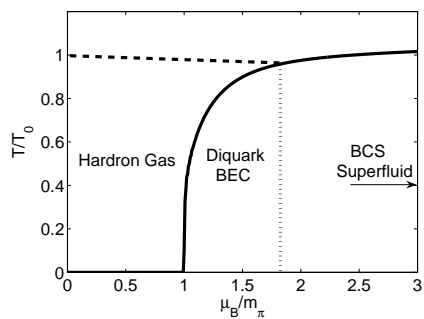}
\caption{The proposed phase diagram of diquark condensation in two
color QCD in the $T-\mu_B$ plane. \label{fig9}}
\end{center}
\end{figure}

We can now summarize the above results and propose a phase diagram
of two color QCD in the $T-\mu_B$ plane. The phase diagram is
shown in Fig.\ref{fig9}. At low temperature and low baryon
chemical potential, the matter should be in the normal hadron
state. The thick dashed line at high temperature means the
estimated phase transition from hadron gas to quark gas, which can
not be calculated in the NJL model but should exist in the system.
When the baryon chemical potential becomes larger than the pion
mass in the vacuum, the diquark BEC appears and exists up to
another critical baryon chemical potential $\mu_B^0$ indicated by
the vertical dashed line. At high enough baryon chemical
potential, the matter will become the BCS superfluid where the
quark-quark Cooper pairs are condensed. Between the BEC and BCS
states there should exist a crossover region, like the pseudogap
regime in high temperature superconductor and ultracold fermion
gas. Note that the transition from diquark BEC to BCS superfluid
is a smooth crossover. Since two color QCD can be successfully
simulated on lattice, such a BEC-BCS crossover can be confirmed by
measuring the quark energy gap and comparing it with the diquark
condensate. The pseudogap phase at high temperature can also be
confirmed by investigating the quark spectral function.

\section {Diquark BEC-BCS Crossover in Three Color NJL Model}
\label{s3}
Let us now consider the realistic world with color degrees of
freedom $N_c=3$. The three color NJL model including the scalar
diquark channel is defined as
\begin{eqnarray}
{\cal L} &=&
\bar{\psi}\left(i\gamma^{\mu}\partial_{\mu}-m_0\right)\psi
+G_s\left[\left(\bar{\psi}\psi\right)^2+\left(\bar{\psi}i\gamma_5
\tau\psi\right)^2 \right]\nonumber\\
&&+G_d\sum_{a=2,5,7}\left(\bar\psi i\gamma^5\tau_2\lambda_a
C\bar{\psi}^{\text T}\right)\left(\psi^{\text T}C
i\gamma^5\tau_2\lambda_a\psi\right),\nonumber\\
\end{eqnarray}
where $\lambda_a$ are the Gell-Mann matrices in color space.
Different from two color QCD, the scalar diquarks in three color
case are color anti-triplets. Therefore, in real QCD, the problem
of confinement becomes important. In this section, we will discuss
the possible diquark BEC-BCS crossover without considering the
effect of confinement.

\subsection {Scalar Diquark Mass in the Vacuum}
We first discuss the scalar diquark mass in the vacuum with
$T=\mu_B=0$. In the random phase approximation, the diquark mass
is determined by the pole equation
\begin{equation}
1-2G_d\Pi_d(k_0=m_d,{\bf k}=0)=0,
\end{equation}
where the diquark polarization function is defined as
\begin{equation}
\Pi_d(k)=4i\int\frac{d^4 p}{(2\pi)^4}\text{Tr}\left[i\gamma_5{\cal
S}(p+k)i\gamma_5{\cal S}(p)\right]
\end{equation}
with the mean field quark propagator ${\cal S}(p)=(\gamma^\mu
p_\mu-m)^{-1}$. The effective quark mass $m$ defined as
$m=m_0-2G_s\langle\bar{\psi}\psi\rangle$ satisfies the gap
equation
\begin{equation}
m-m_0=24G_s m\int{d^3{\bf p}\over (2\pi)^3}{1\over \sqrt{{\bf
p}^2+m^2}}.
\end{equation}

Taking the trace in the Dirac space and summation over the quark
frequency, the diquark mass $m_d$ is controlled by
\begin{equation}
1=8G_d\int{d^3{\bf p}\over (2\pi)^3}\left(\frac{1}{E_{\bf
p}+m_d/2}+\frac{1}{E_{\bf p}-m_d/2}\right).
\end{equation}
If diquarks can exist as stable bound states in the vacuum, their
mass must satisfy the constraint $0<m_d<2m$. As a consequence, the
coupling constant in the scalar diquark channel should be in the
region $G_d^{\text min}<G_d<G_d^{\text{max}}$\cite{ebert,zhuang2},
where the upper limit is nearly model independent,
\begin{equation}
G_d^{\text{max}}=\frac{3}{2}G_s\frac{m}{m-m_0}\simeq\frac{3}{2}G_s,
\end{equation}
but the lower limit depends on the model parameters,
\begin{equation}
G_d^{\text{min}}=\frac{\pi^2}{4\left(\Lambda\sqrt{\Lambda^2+m^2}
+m^2\ln{\frac{\Lambda+\sqrt{\Lambda^2+m^2}}{m}}\right)}.
\end{equation}

In the three color NJL model, the model parameters are set to be
$m_0=5$ MeV, $G_{\text s}=4.93$ GeV$^{-2}$ and $\Lambda=653$ MeV.
For convenience, we take the coupling ratio $\eta=G_d/G_s$ instead
of $G_d$. With the above parameter values, we find
$\eta_{\text{max}}\simeq 1.55$ and $\eta_{\text{min}}\simeq 0.82$.
For other possible parameter values, the lower limit is roughly in
the region $0.7<\eta_{\text{min}}<0.8$. Physically speaking, if
$\eta>\eta_{\text{max}}$, the vacuum becomes unstable, and if
$\eta<\eta_{\text{min}}$, a scalar diquark can decay into two
quarks and becomes a unstable resonance.

\begin{figure}[!htb]
\begin{center}
\includegraphics[width=6cm]{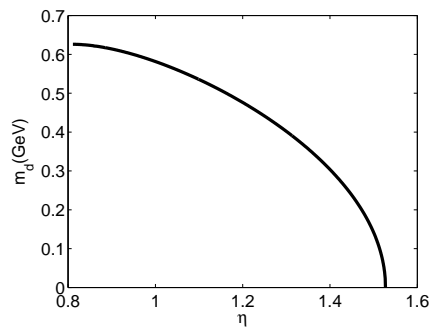}
\caption{The scalar diquark mass $m_d$ in the vacuum as a function
of the coupling ratio $\eta=G_d/G_s$. \label{fig10}}
\end{center}
\end{figure}

The $\eta$ dependence of the scalar diquark mass is shown in
Fig.\ref{fig10}. As estimated in \cite{rapp}, a scalar diquark is
in a deeply bound state when the binding energy is in the region
$200$ MeV $<2m-m_d<300$ MeV, which corresponds to the diquark mass
$300$ MeV $<m_d<400$ MeV and to the coupling ratio $1.3<\eta<1.4$.

\subsection {Diquark BEC-BCS crossover}
At large enough baryon chemical potential, the diquarks will
condense. The diquark condensate is defined as
\begin{equation}
\Delta_a=-2G_d\langle\psi^{\text T}C
i\gamma^5\tau_2\lambda_a\psi\rangle.
\end{equation}
Due to the color SU(3) symmetry, we can choose a specific gauge
$\Delta_2=\Delta\neq 0, \Delta_5=\Delta_7=0$. In this gauge, the
red and green quarks participate in the condensation, but the blue
one does not. We firstly consider the simplest case where the
chemical potentials for the red, green and blue quarks are equal,
$\mu_r=\mu_g=\mu_b=\mu_B/3$. General cases with color neutrality
and vector meson coupling will be discussed in the following
subsections. The thermodynamic potential at mean field level
reads\cite{huang}
\begin{eqnarray}
\Omega&=&\frac{(m-m_0)^2}{4G_s}+\frac{\Delta^2}{4G_d}\\
&&-4\int {d^3 {\bf p}\over (2\pi)^3}\left[2\left(\zeta(E_\Delta^+)
+\zeta(E_\Delta^-)\right)+\zeta(E_b^+)+\zeta(E_b^-)\right],\nonumber
\end{eqnarray}
where the quark energies are defined as
\begin{eqnarray}
&&E_\Delta^\pm=\sqrt{\left(E_{\bf p}\pm\mu_B/3\right)^2+\Delta^2},\nonumber\\
&&E_{\text b}^\pm=E_{\bf p}\pm\mu_B/3.
\end{eqnarray}
Minimizing the thermodynamic potential, we obtain the gap
equations for the chiral and color condensates $m$ and $\Delta$ at
zero temperature,
\begin{eqnarray}
m-m_0&=&8G_s m\int{d^3{\bf p}\over (2\pi)^3}{1\over E_{\bf p}}
\left[{E_{\bf p}^-\over E_\Delta^-}+{E_{\bf p}^+\over
E_\Delta^+}+\Theta\left(E_b^-\right)\right],\nonumber\\
\Delta&=&8G_d\Delta\int{d^3{\bf p}\over (2\pi)^3}\left({1\over
E_\Delta^-}+{1\over E_\Delta^+}\right).
\end{eqnarray}

The above gap equations are almost the same as in the two color
case, except the term $\Theta(E_b^-)$ in the first equation. For
$\eta<\eta_{\text{min}}$, diquark condensation takes place at
about $\mu_B^c\simeq 3m$ and the phase transition is of first
order. In this case, there exists no BEC region where
$\mu_N\equiv\mu_r-m=\mu_B/3-m$ is negative. On the other hand, For
$\eta_{\text{min}}<\eta<\eta_{\text{max}}$, the diquarks become
condensed at $\mu_B^c=3m_d/2$ and the phase transition is of
second order. The proof is as follows. For $\mu_B<3m_d/2<3m$, the
gap equation for $m$ keeps the same form as in the vacuum, and the
gap equation for $\Delta$ becomes
\begin{equation}
1=8G_d\int{d^3{\bf p}\over (2\pi)^3}\left(\frac{1}{E_{\bf
p}+\mu_B^c/3}+\frac{1}{E_{\bf p}-\mu_B^c/3}\right)
\end{equation}
at the phase transition of color superconductivity with
$\Delta=0$. From the comparison with the diquark mass equation in
the vacuum, the critical baryon chemical potential for color
superconductivity should be $\mu_B^c=3m_d/2$\cite{zhuang2}. In
this case, there must exist a BEC region where $\mu_N$ is
negative, since at the phase transition point we have
$\mu_N=\mu_B^c/3-m=m_d/2-m<0$.

\begin{figure}[!htb]
\begin{center}
\includegraphics[width=6cm]{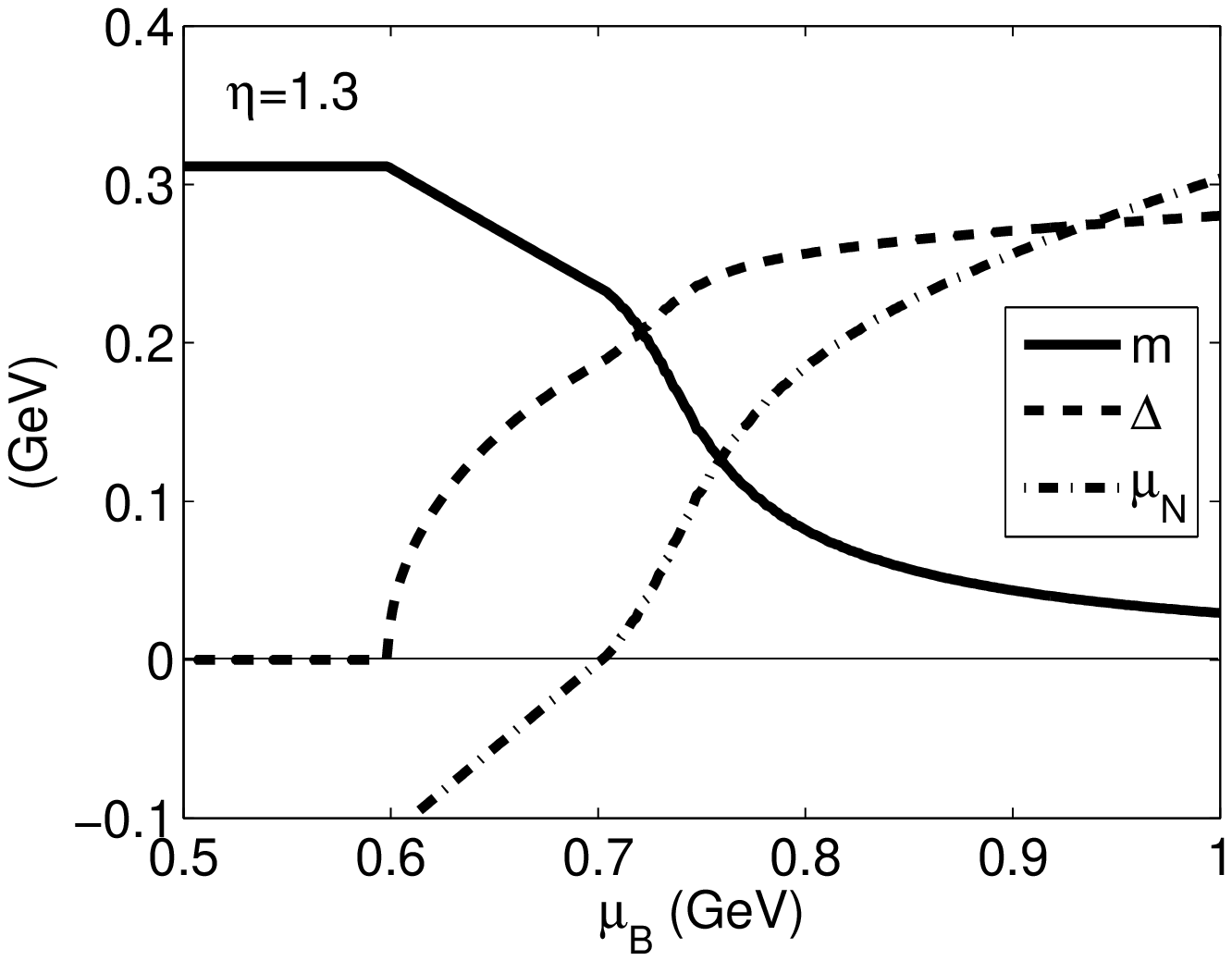}
\includegraphics[width=6cm]{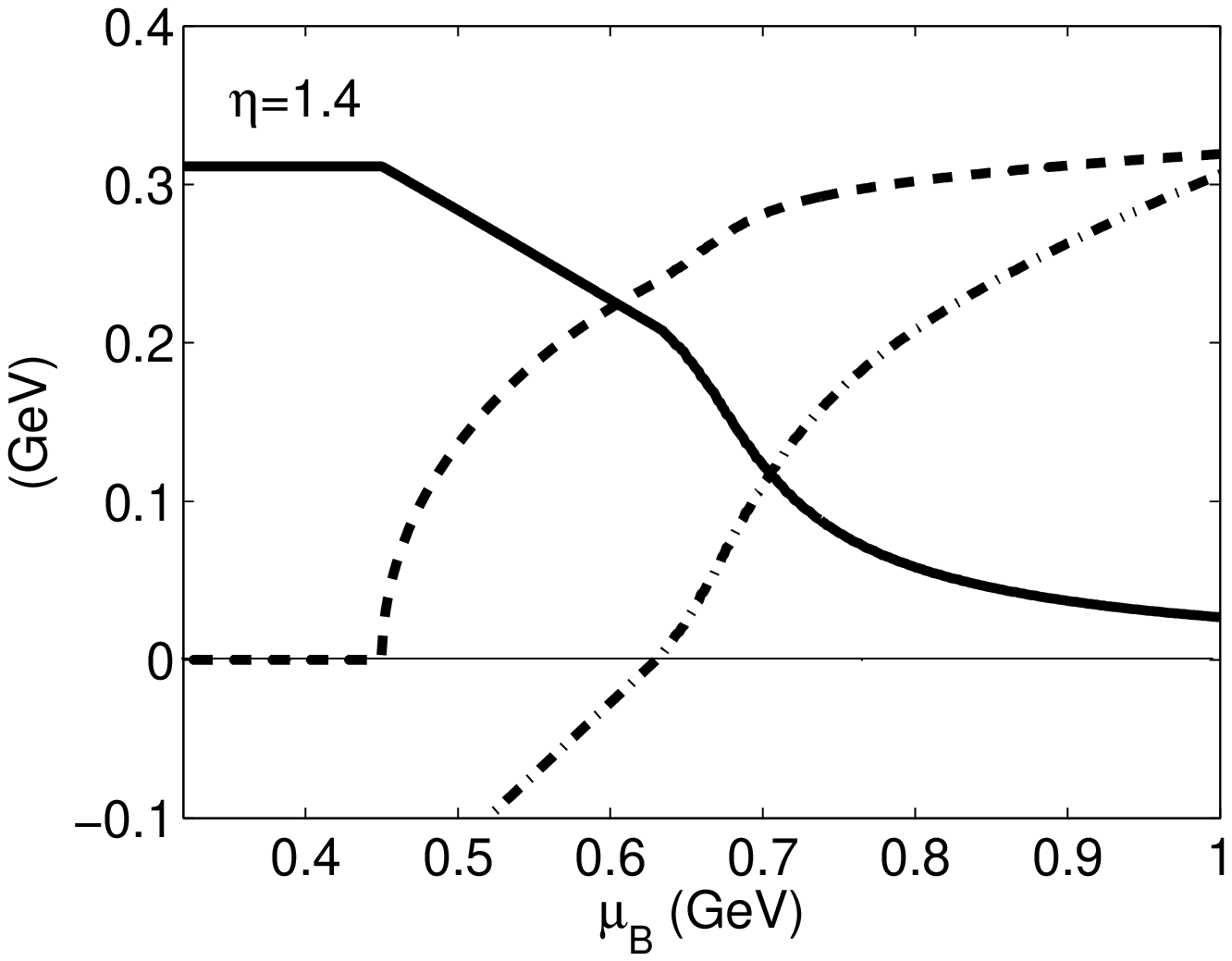}
\includegraphics[width=6cm]{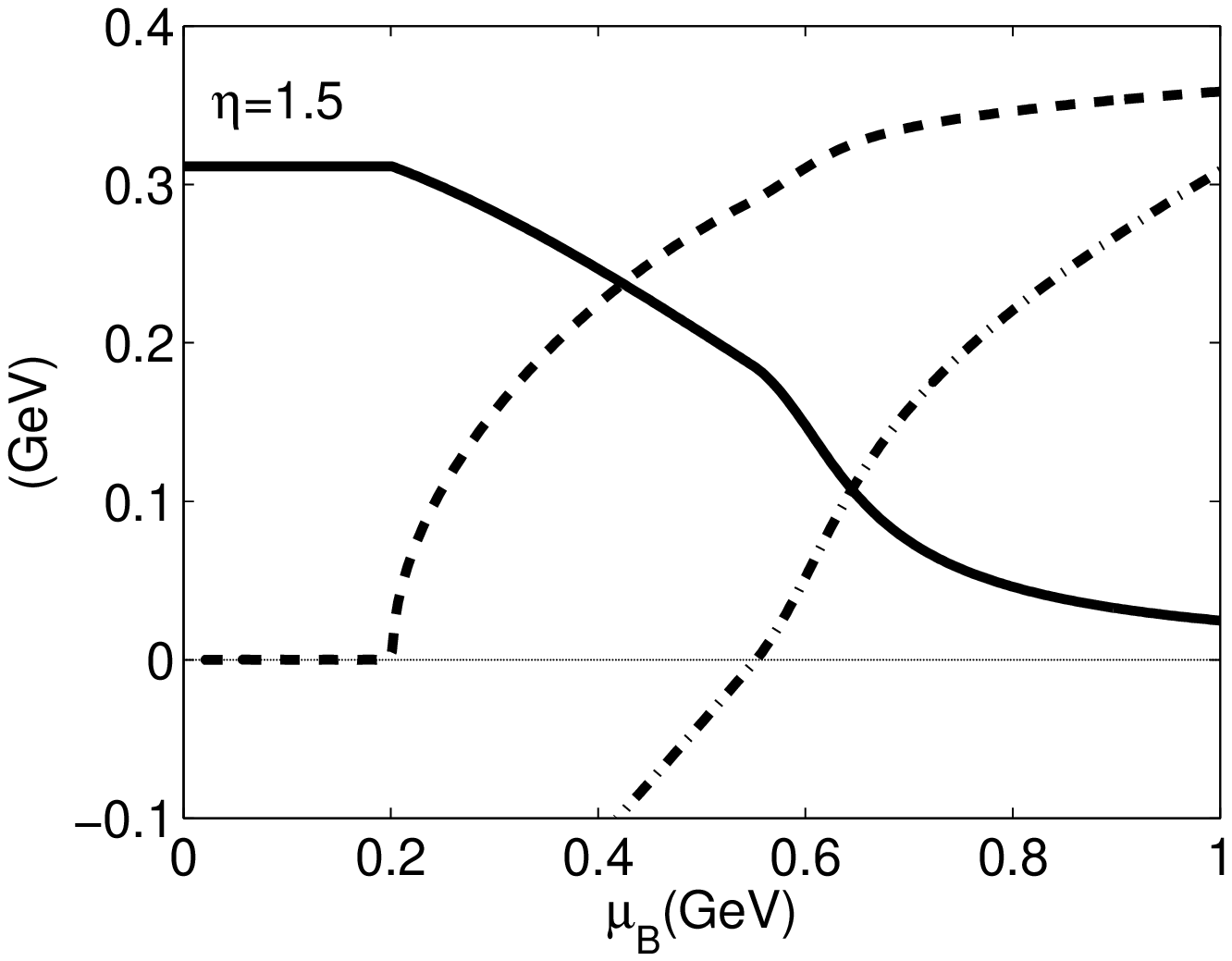}
\caption{The diquark condensate $\Delta$, effective quark mass $m$
and the chemical potential $\mu_N$ as functions of baryon chemical
potential for several values of the coupling ratio $\eta$ at zero
temperature. The color neutrality is not taken into account.
\label{fig11}}
\end{center}
\end{figure}

In Fig.\ref{fig11}, we show the effective quark mass and diquark
condensate as functions of baryon chemical potential. From the
behavior of $\mu_N$, there exists indeed a diquark BEC region at
intermediate baryon chemical potential. At sufficiently large
baryon chemical potential the diquark condensate is in the BCS
form.

In the diquark BEC region, we assume that the effective quark mass
behaves as
\begin{equation}
\frac{m(\mu_B)}{m(0)}\simeq\left(\frac{\mu_B^c}{\mu_B}\right)^\alpha,\
\ \ \alpha>0.
\end{equation}
From our numerical calculation, the value of $\alpha$ depends on
the coupling $\eta$. Taking $m(\mu_B)=\mu_B/3$ from $\mu_N=0$ at
the end point of the BEC, the diquark BEC region ends at
\begin{equation}
\mu_B^0=\frac{3}{2}\left[2m(0)m_d^\alpha\right]^{\frac{1}{\alpha+1}}.
\end{equation}
Therefore, the interval of $\mu_{\text B}$ for diquark BEC is
\begin{equation}
\Delta\mu_B=\mu_B^0-\mu_B^c=\frac{3}{2}m_d\left[\left(\frac{2m(0)}{m_d}\right)^{\frac{1}{\alpha+1}}-1\right].
\end{equation}
The largest BEC region takes place at $m_{\text d}\simeq 185$ MeV
corresponding to $\eta\simeq1.46$.

The critical temperature $T_c$ can be determined by solving the
gap equations with $\Delta=0$. Above the critical temperature, all
scalar diquarks become degenerate and their polarization function
reads $\Pi_d(k)=\Pi_3(k;2\mu_B/3)$ with $\Pi_3$ listed in Appendix
\ref{app1}. Similar to the calculation in the two color case, the
diquarks are in bound state in the BEC region with $\mu_B/3<m$,
and become unstable resonances in the BCS region. The numerical
results and discussions are quite similar to that in the two color
NJL model.

\subsection {Effect of Color Neutrality}
In three color QCD,  diquarks are no longer colorless and the
requirement of color neutrality is not automatically satisfied
once diquarks are included. This can be seen if we compute the
expectation values of the color charges $\langle
Q_a\rangle=\langle \bar{\psi}\lambda_a\gamma_0\psi\rangle$. In the
gauge with $\Delta_2\neq 0$ and $\Delta_5=\Delta_7=0$, we find
$\langle Q_8\rangle\neq0$. To solve this problem, we can introduce
a color chemical potential $\mu_8$ corresponding to the 8-th color
charge. It may be dynamically generated by a gluon condensate
$\langle A_8^0\rangle$. The color chemical potential enters the
gap equations by replacing the quark chemical potentials
$\mu_r=\mu_g=\mu_b=\mu_B/3$ by $\mu_r=\mu_g=\mu_B/3+\mu_8/3$ and
$\mu_b=\mu_B/3-2\mu_8/3$. The color charge neutrality condition is
guaranteed by $n_8=-\partial\Omega/\partial\mu_8=0$, namely
\begin{equation}
\int{d^3{\bf p}\over(2\pi)^3}\left[{E_{\bf p}^+\over
E_\Delta^+}-{E_{\bf p}^-\over
E_\Delta^-}-2\Theta\left(-E_b^-\right)\right]=0.
\end{equation}

\begin{figure}[!htb]
\begin{center}
\includegraphics[width=6cm]{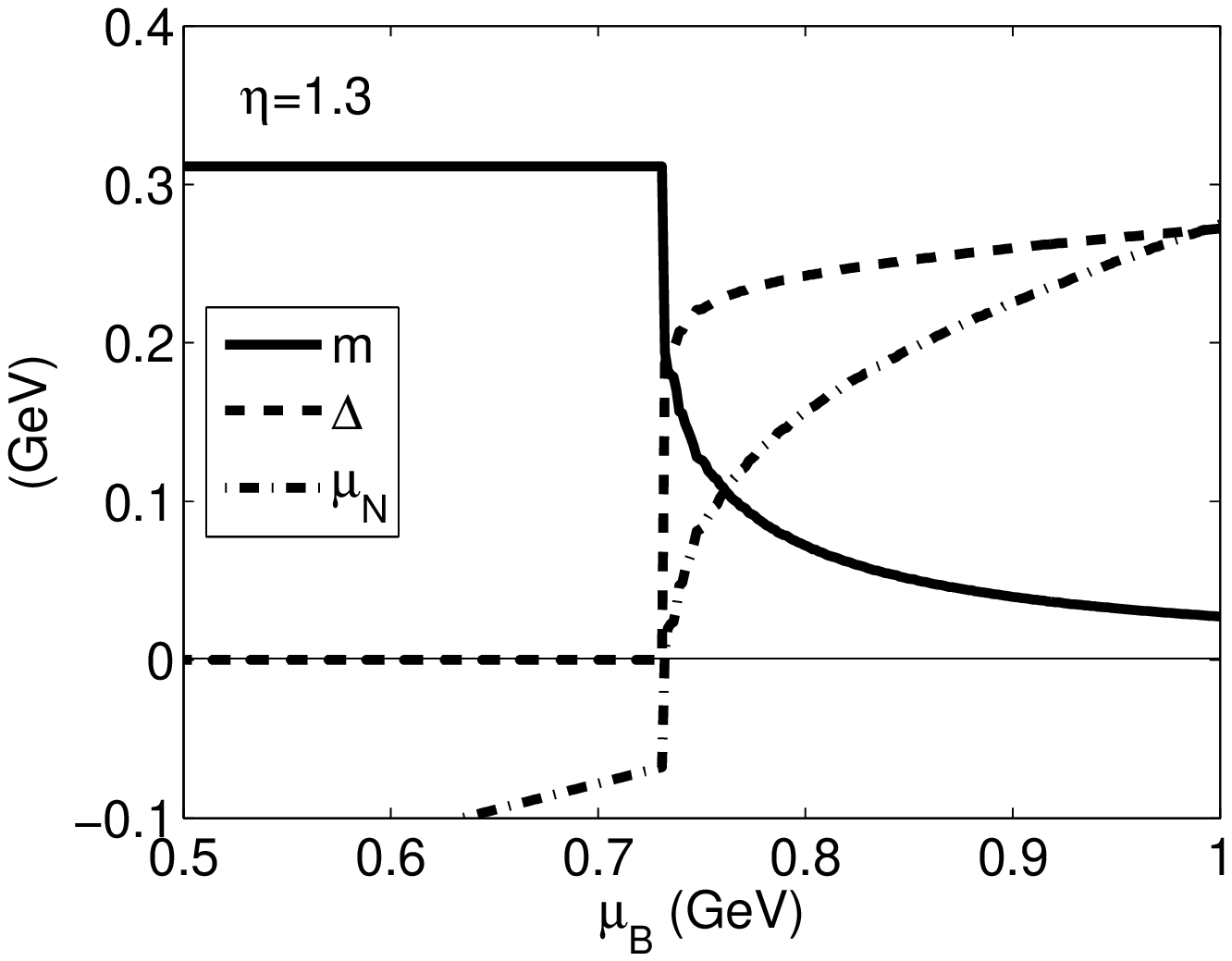}
\includegraphics[width=6cm]{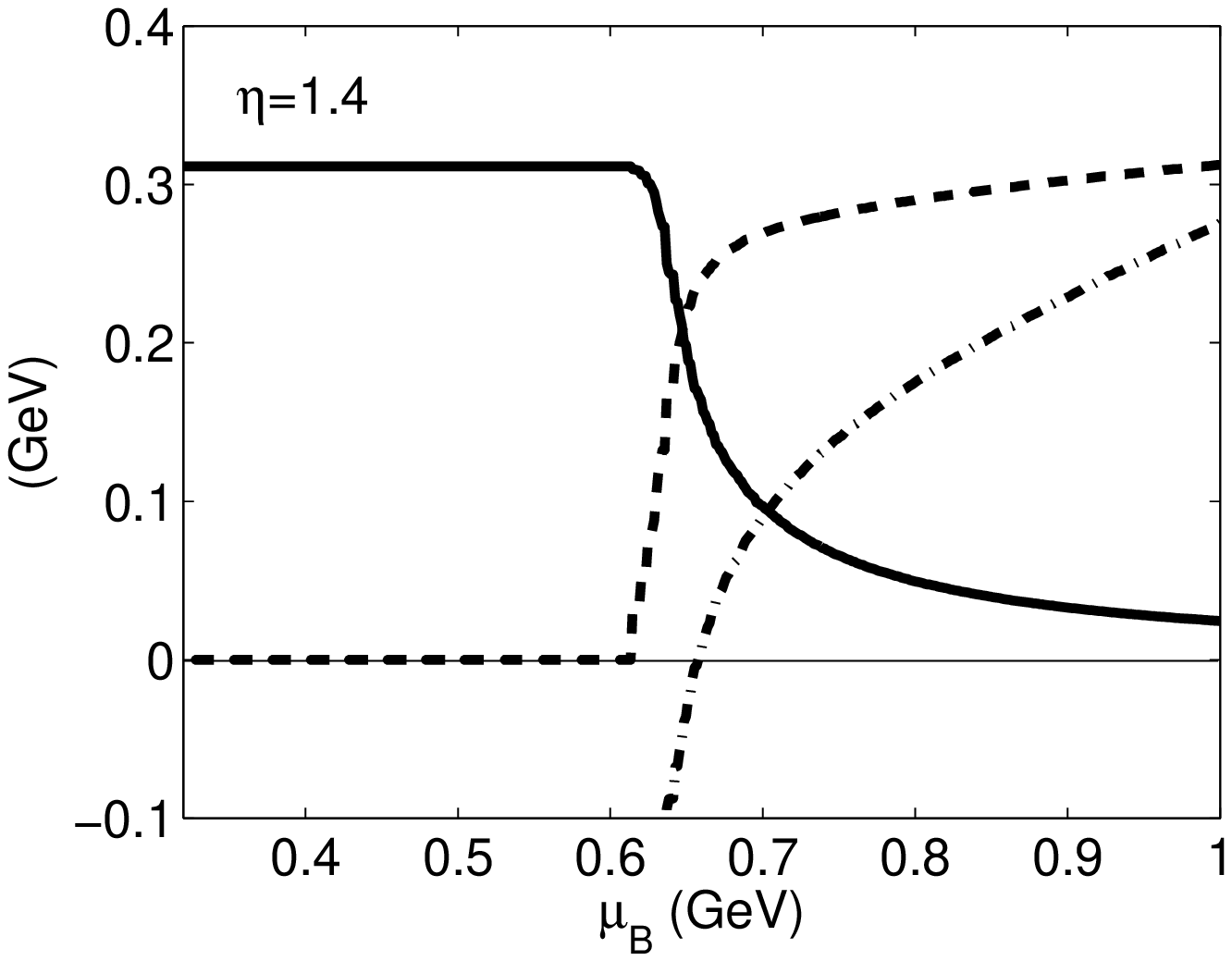}
\includegraphics[width=6cm]{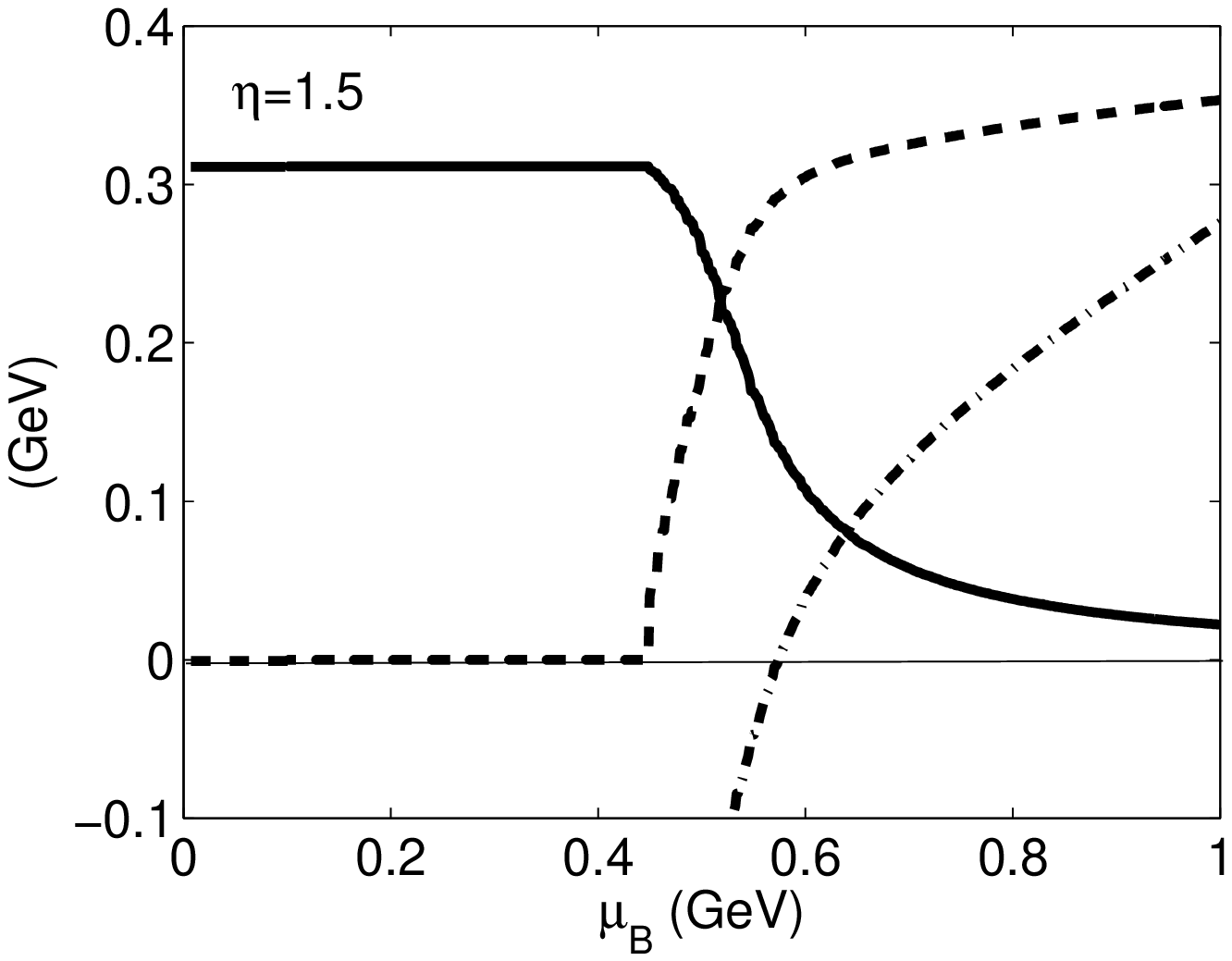}
\caption{The diquark condensate $\Delta$, effective quark mass $m$
and the chemical potential $\mu_N$ as functions of baryon chemical
potential for several values of the coupling ratio $\eta$ at zero
temperature. The color neutrality is considered. \label{fig12}}
\end{center}
\end{figure}

Taking into account the color neutrality, the recalculated
effective quark mass and diquark condensate are shown in
Fig.\ref{fig12}. Comparing the behavior of the effective chemical
potential $\mu_N\equiv\mu_r-m$ in the two cases with and without
considering color neutrality, we find that the requirement of
color neutrality disfavors diquark BEC. However, it does not
cancel the BEC region completely.

\subsection {Effect of Vector Meson Coupling }
Now we ask the question whether there exists some mechanism that
favors the diquark BEC. As we have seen, the diquark BEC happens
in the beginning part of color superconductivity where the broken
chiral symmetry starts to restore and the effective quark mass is
still large enough. As proposed in \cite{kitazawa}, the quark
interaction in vector meson channel may be a candidate to slow
down the chiral symmetry restoration and enhance the BEC
formation. Now we include a new interaction term to the NJL
lagrangian,
\begin{equation}
{\cal L}_{\text v} =-G_{\text
v}\left[\left(\bar{\psi}\gamma_\mu\psi\right)^2+\left(\bar{\psi}\gamma_\mu\gamma_5\tau\psi\right)^2
\right],
\end{equation}
which corresponds to vector mesons. At finite density, a new
condensate $\rho_{\text v}=2G_{\text v}\langle
\bar\psi\gamma_0\psi\rangle$ which is proportional to the baryon
number density should be considered. This condensate induces a new
energy term $-\rho_{\text v}^2/4G_{\text v}$ in the thermodynamic
potential and it enters the gap equations by replacing $\mu_B/3$
by $\mu_B/3-\rho_{\text v}$. The gap equation for $\rho_{\text v}$
at zero temperature reads
\begin{equation}
\rho_{\text v}=8G_{\text v}\int{d^3{\bf
p}\over(2\pi)^3}\left[{E_{\bf p}^+\over E_\Delta^+}-{E_{\bf
p}^-\over E_\Delta^-}+\Theta\left(-E_b^-\right)\right].
\end{equation}

\begin{figure}[!htb]
\begin{center}
\includegraphics[width=6cm]{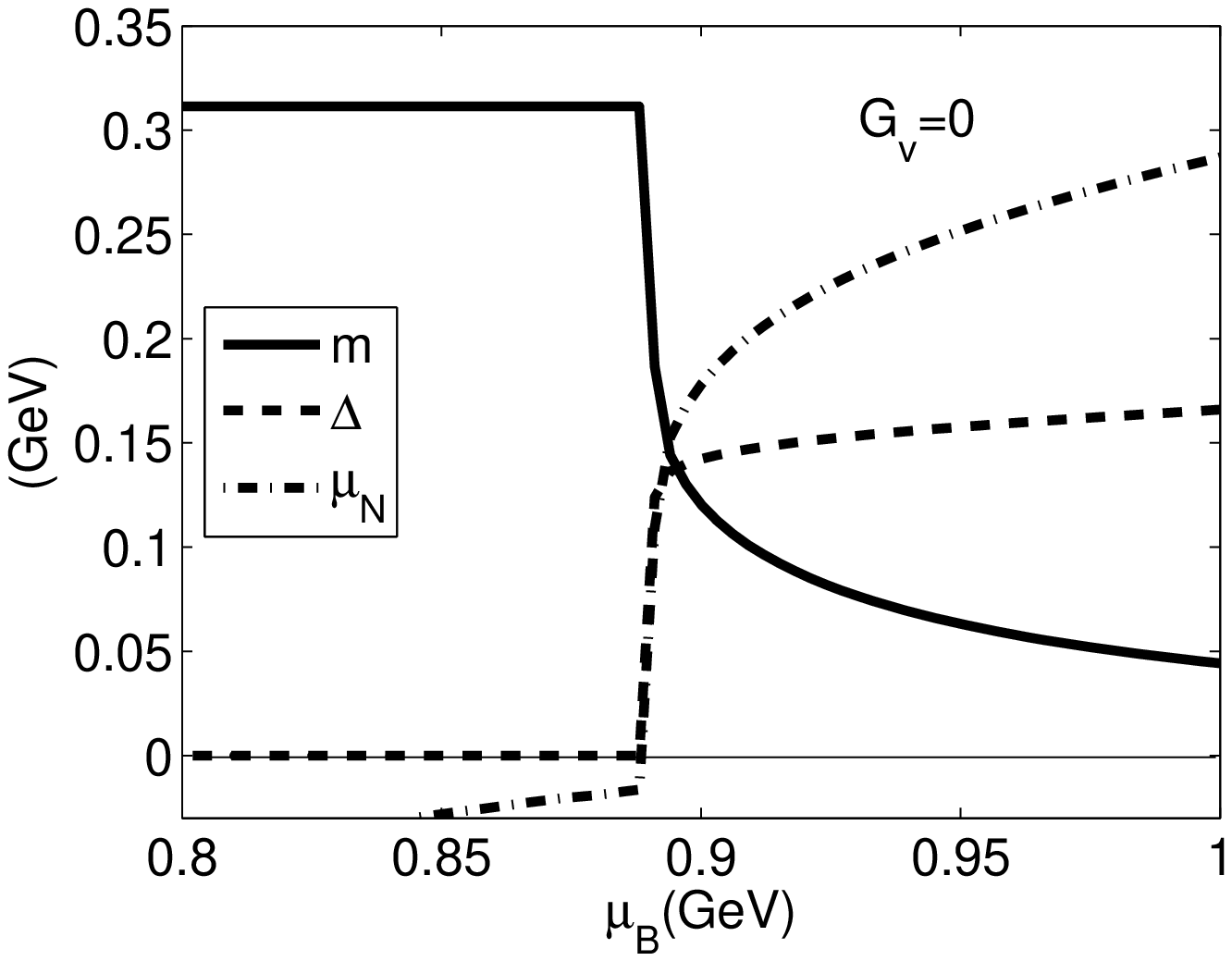}
\includegraphics[width=6cm]{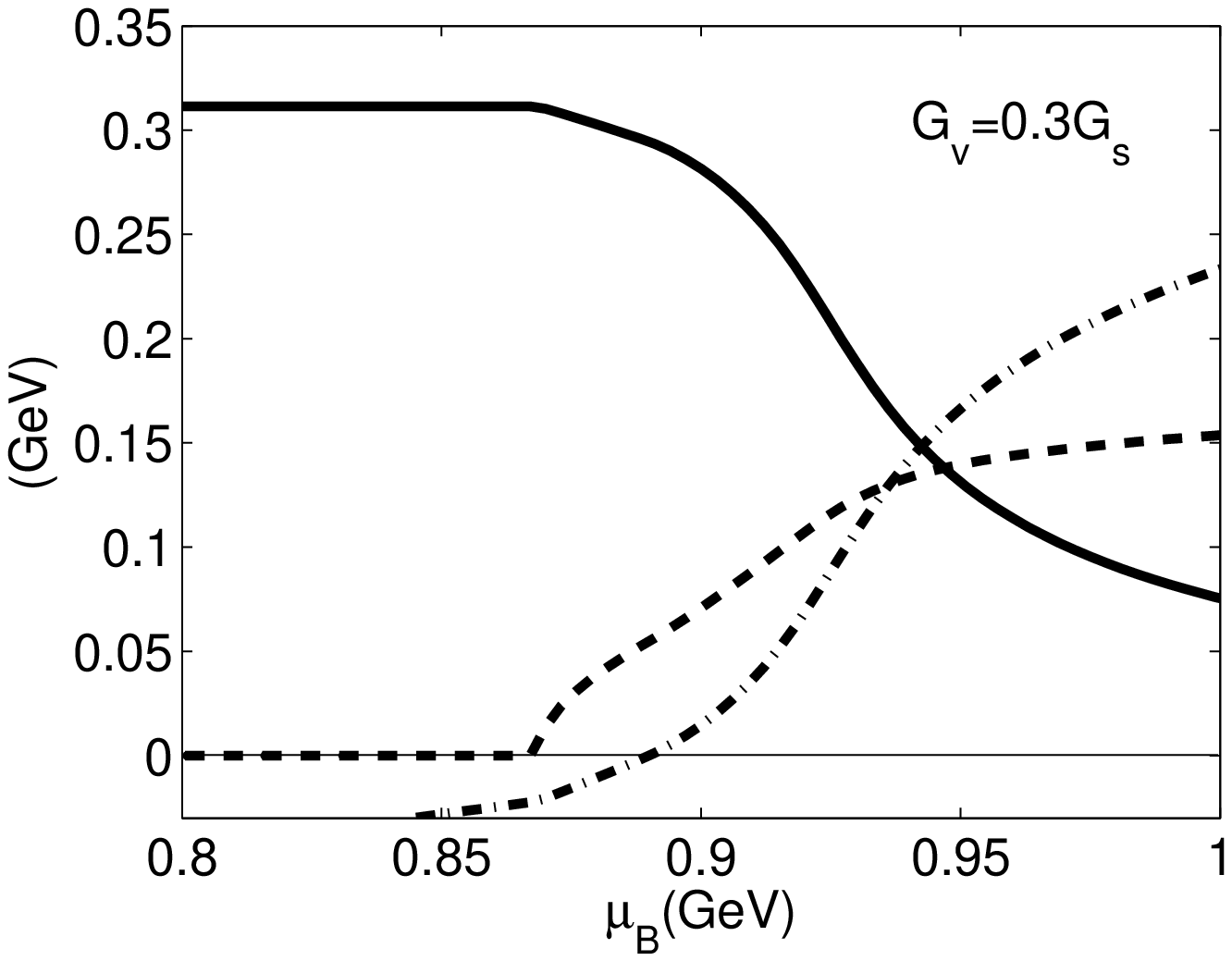}
\includegraphics[width=6cm]{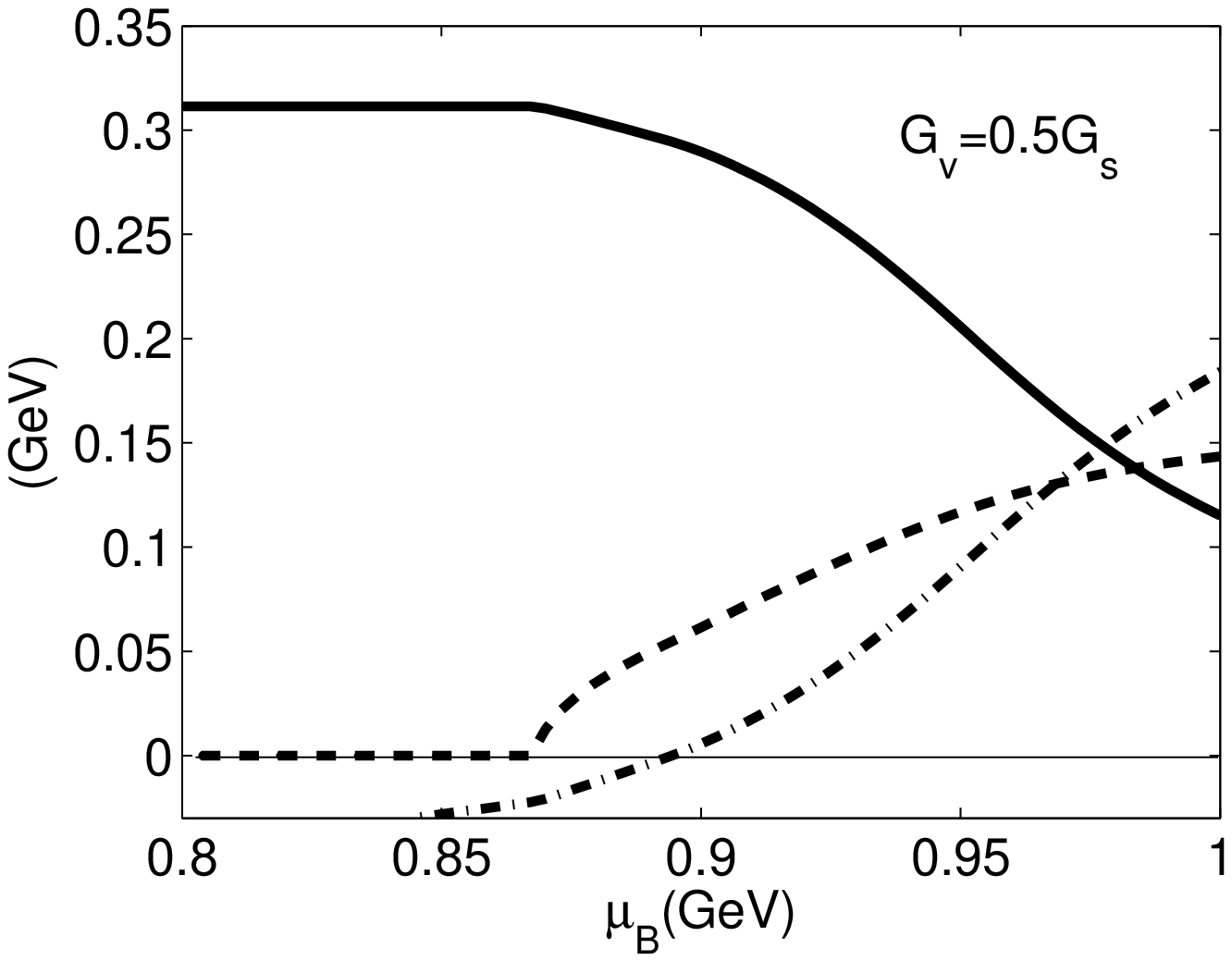}
\caption{The diquark condensate $\Delta$, effective quark mass $m$
and the chemical potential $\mu_N$ as functions of baryon chemical
potential for several values of the vector couplings $G_{\text v}$
and at a fixed diquark coupling ratio $\eta=1$ at zero
temperature. \label{fig13}}
\end{center}
\end{figure}

In Fig.\ref{fig13} we calculate the effective quark mass and the
diquark condensate at the coupling ratio $\eta=1$ and for three
values of the vector coupling $G_{\text v}$. For $G_{\text v}=0$
the diquark BEC region is almost not visible, but for a reasonably
large vector coupling such as $G_{\text v}/G_s=0.3$ or $0.5$,
there appears a large diquark BEC region. It is clear that the
vector meson channel really slows down the chiral symmetry
restoration and favors the formation of diquark BEC. There may
exist other mechanisms that favor BEC, such as the axial anomaly
in QCD\cite{hatsuda2}.

\subsection {Chromomagnetic Instability}
For realistic QCD matter in nature such as in compact stars, beta
equilibrium and charge neutrality should be considered. These
effects induce a chemical potential mismatch $\delta\mu$ between
$u$ and $d$ quarks. Assuming the ground state to be a uniform
phase, the mismatch effect will lead to an interesting gapless
color superconductivity phases\cite{gapless1,gapless2}. However,
it was found that the gapless phase in the weak coupling region
suffers the so-called chromomagnetic instability\cite{CI}: The
Meissner masses squared of some gluons are negative. While the
chromomagnetic instability indicates that the ground state of the
superfluid may favor to be in some non-uniform phase such as LOFF
phase\cite{LOFF1,LOFF2,LOFF3}, the instability may be cured in
some region. Based on a two species model\cite{kitazawa2}, it is
found that the magnetic instability should be fully cured in the
BEC region. However, the 4-7th gluons' instability is not yet
examined. Here we try to complete this work.

\begin{figure}[!htb]
\begin{center}
\includegraphics[width=6cm]{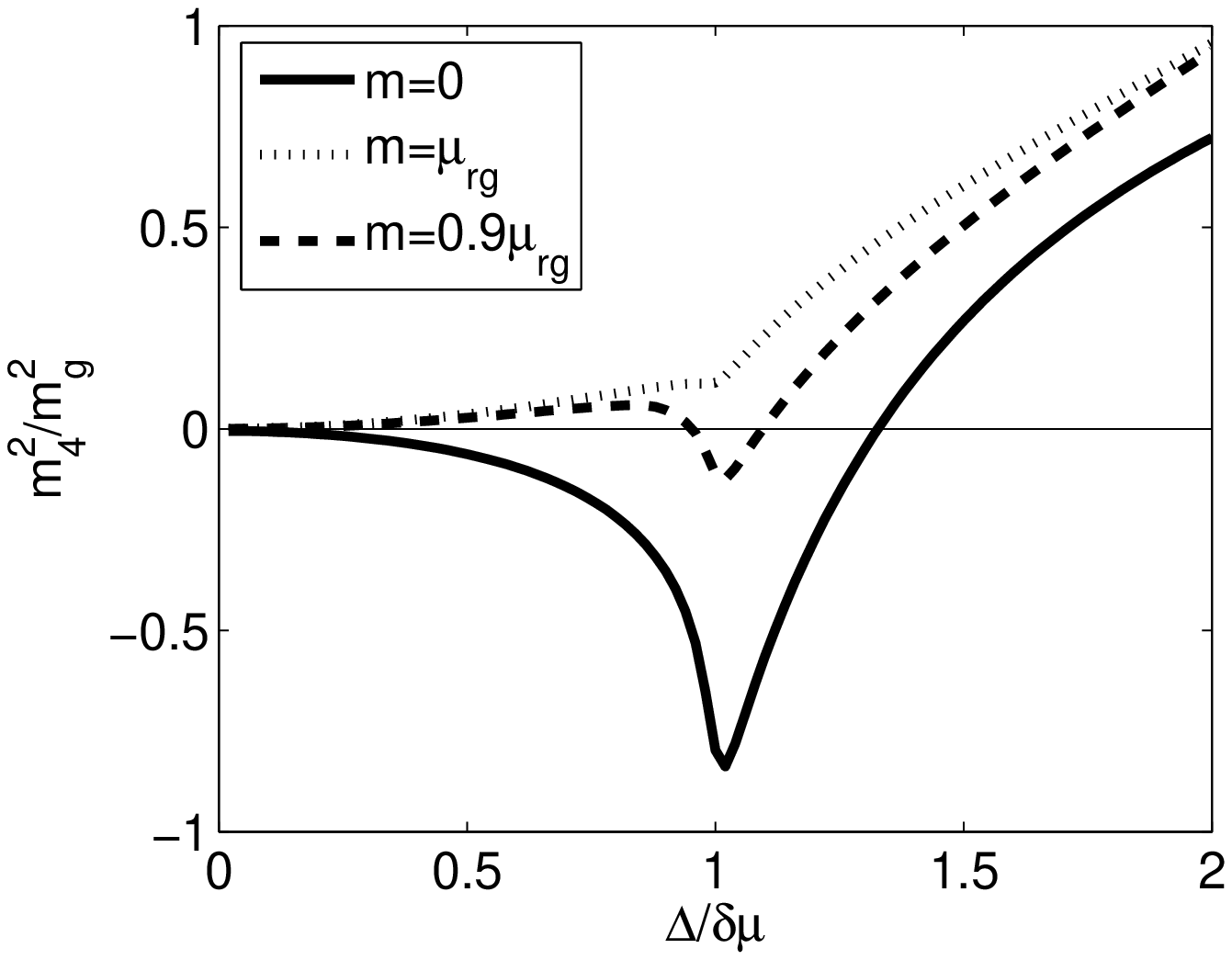}
\includegraphics[width=6cm]{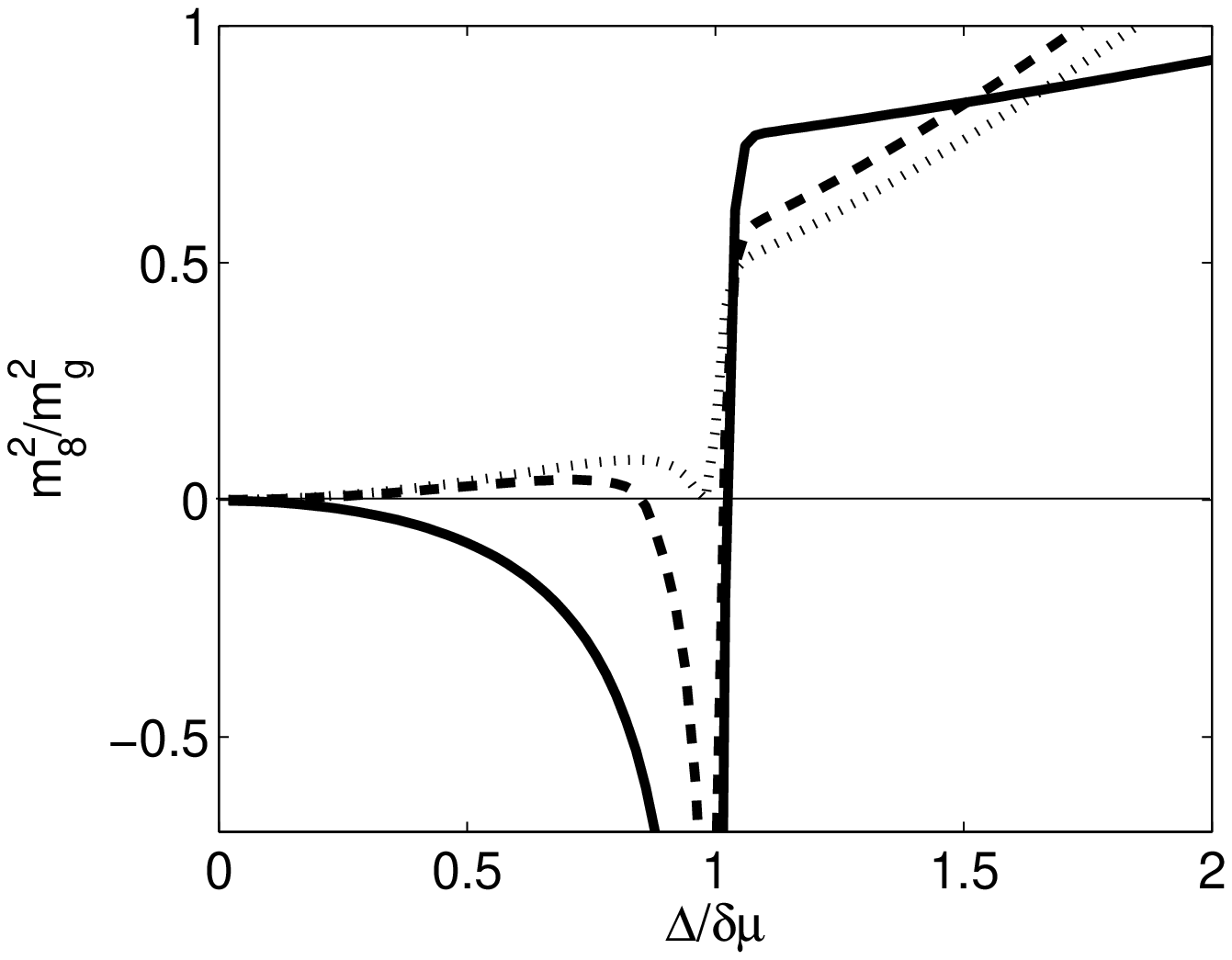}
\caption{The Meissner masses squared, scaled by
$m_g^2=4\alpha_s\mu_r^2/(3\pi)$ with $\alpha_s$ being the QCD
gauge coupling constant, for the 4-7th gluons (upper panel) and
8th gluon (lower panel) as functions of $\Delta/\delta\mu$.
\label{fig14}}
\end{center}
\end{figure}

The analytical expressions of the Meissner masses squared $m_4^2$
for the 4-7th gluons and $m_8^2$ for the 8th gluon are listed in
Appendix \ref{app2}. We show the Meissner masses squared in
Fig.\ref{fig14} in both BCS and BEC states. In the BCS case with
$m\ll\mu_r$, our result is consistent with the analytical
expression\cite{CI}. When we approach to the BEC region, both the
4-7th and 8th gluons' instabilities are partially cured. In the
BEC region with $m\ge\mu_r$, they are fully cured. Our
investigation here is consistent with the study based on a
non-relativistic model\cite{he2}. Since the chromomagnetic
instability is fully cured, the ground state in the BEC region
should be a uniform phase. We should address that such a
phenomenon is quite similar to what happens in non-relativistic
systems\cite{NR1,NR2,NR3}.

\section {Pion BEC-BCS Crossover at Finite Isospin Density}
\label{s4}
Another BEC-BCS crossover in dense QCD happens at finite isospin
density where pions become condensed\cite{son}. At small isospin
density, the QCD matter is a BEC of charged pions, but at ultra
high isospin density, deconfinement happens and the matter turns
to be a BCS superfluid with quark-antiquark cooper
pairing\cite{son}. Therefore, there should be a BEC to BCS
crossover when the isospin chemical potential increases. From the
similarity between pions in three color QCD and diquarks in two
color QCD, the pion superfluidity discussed in this section is
quite similar to the diquark superfluid in Section \ref{s2}. We
start from the two flavor Nambu--Jona-Lasinio model with only
scalar meson channel,
\begin{equation}
{\cal L} =
\bar{\psi}\left(i\gamma^{\mu}\partial_{\mu}-m_0\right)\psi
+G_s\left[\left(\bar{\psi}\psi\right)^2+\left(\bar{\psi}i\gamma_5\tau\psi\right)^2
\right].
\end{equation}
The key quantity describing the system is the partition function
\begin{equation}
Z=\int[d\bar{\psi}][d\psi]e^{\int_0^\beta d\tau\int d^3{\bf
x}\left[{\cal
L}+\frac{\mu_I}{2}\bar{\psi}\gamma_0\tau_3\psi\right]},
\end{equation}
where the isospin chemical potential $\mu_I$ corresponds to the
third component $I_3$ of the isospin charge. Using the
Hubbard-Stratonovich transformation we introduce the auxiliary
meson fields $\sigma$ and $\pi$, and the partition function can be
written as
\begin{equation}
Z=\int[d\bar{\psi}][d\psi][d\sigma][d\pi]e^{\int_0^\beta d\tau\int
d^3{\bf x}{\cal L}_{\text{eff}}}
\end{equation}
with the effective Lagrangian
\begin{equation}
{\cal L}_{\text{eff}}=\bar{\psi}{\cal
K}[\sigma,\pi]\psi-\frac{\sigma^2+\pi^2}{4G_s}
\end{equation}
where the kernel ${\cal K}$ is defined as
\begin{equation}
{\cal K}[\sigma,\pi]= i\gamma^\mu\partial_\mu-m_0+{\mu_I\over
2}\gamma_0\tau_3-(\sigma+i\gamma_5\tau\cdot\pi).
\end{equation}
Integrating out the quark degrees of freedom, we obtain
\begin{equation}
Z=\int[d\sigma][d\pi]e^{-S_{\text{eff}}[\sigma,\pi]}
\end{equation}
with the meson effective action
\begin{equation}
S_{\text{eff}}[\sigma,\pi]=\int_0^\beta d\tau\int d^3{\bf
x}\frac{\sigma^2+\pi^2}{4G_s}-\text{Tr}\ln{\cal K}[\sigma,\pi].
\end{equation}

\subsection {Pion Fluctuation at $T>T_c$}
As we expected, when the isospin chemical potential $\mu_I$
becomes larger than the pion mass $m_\pi$ in the vacuum, the
charged pions will condense at low temperature. In this subsection
we focus on the region above the critical temperature $T_c$ where
the pion condensate vanishes. After the field shift for $\sigma$,
the effective action at zeroth order in meson fields gives the
mean field thermodynamic potential $\Omega$,
\begin{equation}
\Omega=\frac{1}{\beta
V}S_{\text{eff}}^{(0)}=\frac{(m-m_0)^2}{4G_s}+\frac{1}{\beta
V}\ln\det{\cal S}^{-1},
\end{equation}
where ${\cal S}$ is the quark propagator at mean field level
\begin{equation}
{\cal S}= \left(\gamma^\mu
\partial_\mu-m+{\mu_I\over2}\gamma_0\tau_3\right)^{-1}.
\end{equation}
The quadratic term of the effective action reads
\begin{eqnarray}
S_{\text{eff}}^{(2)}[\sigma,\pi]&=&\int_0^\beta d\tau\int
d^3{\bf x}\frac{\sigma^2+\pi^2}{4G_s}\nonumber\\
&&+\frac{1}{2}\text{Tr}\left\{{\cal S}\Sigma[\sigma,\pi]{\cal
S}\Sigma[\sigma,\pi]\right\},
\end{eqnarray}
where $\Sigma$ is defined as
$\Sigma[\sigma,\pi]=\sigma+i\gamma_5\tau\cdot\pi$. Going to
momentum space, the effective action can be evaluated as
\begin{eqnarray}
S_{\text{eff}}^{(2)}[\sigma,\pi]&=&\frac{1}{2}\sum_k\Bigg[\left[\frac{1}{2G_s}-\Pi_\sigma(k)\right]\sigma(-k)\sigma(k)\nonumber\\
&&+\left[\frac{1}{2G_s}-\Pi_{\pi_0}(k)\right]\pi_0(-k)\pi_0(k)\nonumber\\
&&+\left[\frac{1}{2G_s}-\Pi_{\pi_+}(k)\right]\pi_+(-k)\pi_-(k)\nonumber\\
&&+\left[\frac{1}{2G_s}-\Pi_{\pi_-}(k)\right]\pi_-(-k)\pi_+(k),
\end{eqnarray}
where $\pi_\pm=(\pi_1\pm i\pi_2)/\sqrt{2}$ are the positively and
negatively charged pion fields, and the polarization functions for
the mesons can be evaluated as
\begin{eqnarray}\label{mass}
\Pi_{\sigma}(k)=N_c\Pi_1(k;\mu_I),\ \
\Pi_{\pi_0}(k)=N_c\Pi_2(k;\mu_I),\nonumber\\
\Pi_{\pi_+}(k)=N_c\Pi_3(k;\mu_I),\ \
\Pi_{\pi_-}(k)=N_c\Pi_4(k;\mu_I)
\end{eqnarray}
where $N_c$ is the color degree of freedom.

Without loss of generality, we consider the case
$\mu_{\text{I}}>0$ where the $\pi_+$ mesons become condensed at
low temperature. The transition temperature $T_c$ is determined by
the well-known Thouless criterion
\begin{equation}
1-2G_s\Pi_{\pi_+}(k_0=0,{\bf k}=0)\Big|_{T=T_c}=0
\end{equation}
together with the gap equation for the effective quark mass $m$
derived from the first order derivative of the thermodynamic
potential with respect to $m$,
\begin{equation}
\frac{m-m_0}{8N_cG_sm}=\int{d^3{\bf p}\over
(2\pi)^3}{1-f\left(E_{\bf p}^+\right)-f\left(E_{\bf
p}^-\right)\over E_{\bf p}} \Bigg|_{T=T_c}
\end{equation}
with $E_{\bf p}^\pm=E_{\bf p}\pm\mu_I/2$. The critical temperature
as a function of isospin chemical potential is exactly the same as
in Fig.\ref{fig1} if we replace the baryon chemical potential
$\mu_B$ by the isospin chemical potential $\mu_I$. The diquark
condensed phase starts at $\mu_I=m_\pi$, and the critical
temperature can be well described by\cite{splittorff}
\begin{equation}
T_c=T_0\sqrt{1-\left(\frac{m_\pi}{\mu_{\text{I}}}\right)^4},
\end{equation}
where $T_0$ is again the temperature of chiral symmetry
restoration at $\mu_I=0$.

The energy dispersions of the mesons are defined by the poles of
their propagators,
\begin{equation}
1-2G_s\Pi_i(k_0=\omega({\bf k}),{\bf k})=0,\ \ \
i=\sigma,\pi_0,\pi_+,\pi_-.
\end{equation}
Since the sigma and neutral pion do not carry isospin charge, they
do not obtain an isospin chemical potential and their masses are
defined as $m_\sigma=\omega_\sigma(0)$ and
$m_{\pi_0}=\omega_{\pi_0}(0)$. However, for masses of charged
pions, we should subtract the corresponding isospin chemical
potential from the dispersions, namely we take
$m_{\pi_+}=\omega_{\pi_+}(0)+\mu_I$ and
$m_{\pi_-}=\omega_{\pi_-}(0)-\mu_I$. For $\mu_{\text{I}}>0$, at
the transition temperature $T_c$, the $\pi_+$ mass is equal to the
isospin chemical potential,
\begin{equation}
m_{\pi_+}(T_c)=\mu_I.
\end{equation}
The numerical results for meson masses and dissociation
temperatures are the same as in Figs.\ref{fig2} and \ref{fig3} if
we take the correspondence $d\leftrightarrow\pi_+$, $\bar
d\leftrightarrow\pi_-$, $\pi\leftrightarrow \pi_0$ and
$\sigma\leftrightarrow\sigma$ between the two cases.

Similarly, we can investigate the spectral function
$\rho(\omega,{\bf k})$ for $\pi_+$,
\begin{equation}
\rho(\omega,{\bf k})=-2\text{Im}{\text D}_R(\omega,{\bf k}),
\end{equation}
where ${\text D}_R(\omega,{\bf k})\equiv {\cal
D}(\omega+i\eta,{\bf k})$ is the analytical continuation of the
pion Green function ${\cal D}(i\omega_n,{\bf k})$ defined as
\begin{equation}
{\cal D}(i\omega_n,{\bf
k})=\frac{2G_s}{1-2G_s\Pi_{\pi_+}(i\omega_n,{\bf k})}.
\end{equation}
The spectral function is also similar to that of diquarks in
Section \ref{s1}. For $\mu_I<2m$, $\pi_+$ can decay into a
quark-antiquark pair only when its energy satisfies
$\omega>2m-\mu_I$, otherwise it remains a stable bound state.
However, for $\mu_I>2m$, it can decay into a quark-antiquark pair
at any energy and hence becomes an unstable resonance. The
numerical result for the spectral function is similar to that
shown in Fig.\ref{fig4}. In conclusion, with increasing isospin
chemical potential, the pions above the critical temperature
become more and more unstable and finally disappear, which
indicates a BEC-BCS crossover.

We can also define an effective non-relativistic chemical
potential $\mu_N=\mu_I/2-m$. In the deep BCS limit with
$\mu_I\rightarrow\infty$, we have $m\rightarrow 0$ and
$\mu_N\rightarrow\mu_I/2$. In the deep BEC limit with
$\mu_I\rightarrow m_\pi$, $m$ tends to be its vacuum value, and
$\mu_N$ becomes negative and its absolute value approaches to half
of the pion binding energy $2m-m_\pi$. The effective quark mass
$m$ and the chemical potential $\mu_N$ at the critical temperature
are the same as in Fig.\ref{fig5}, and the isospin chemical
potential $\mu_I^0$ for the crossover is still about $240$ MeV.

\subsection { Pion Condensation at $T<T_c$}
At low enough temperature, the pions become condensed, we should
introduce not only the chiral condensate $\langle\sigma\rangle$
but also the pion condensates,
\begin{equation}
\langle\pi_+ \rangle = {\Delta_\pi\over \sqrt 2}e^{i\theta},\ \ \
\ \langle\pi_- \rangle = {\Delta_\pi\over \sqrt 2}e^{-i\theta}.
\end{equation}
A nonzero pion condensate $\Delta_\pi\neq 0$ means spontaneous
breaking of U$_3(1)$ symmetry corresponding to the isospin charge
generator $I_3$. The phase factor $\theta$ indicates the direction
of the U$_3(1)$ symmetry breaking. For a homogeneous superfluid,
we can choose $\theta=0$ without lose of generality. A gapless
Goldstone boson will appear, which can be identified as the
quantum fluctuation in the phase direction.

After a field shift $\sigma\rightarrow
\langle\sigma\rangle+\sigma$ and $\pi_1\rightarrow
\Delta_\pi+\pi_1$, the effective action at zeroth order in meson
fields gives the mean field thermodynamic potential,
\begin{equation}
\Omega=\frac{1}{\beta
V}S_{\text{eff}}^{(0)}=\frac{(m-m_0)^2+\Delta_\pi^2}{4G_{\text
s}}+\frac{1}{\beta V}\ln\det{\cal S}_\pi^{-1},
\end{equation}
where ${\cal S}_\pi$ is the quark propagator in the pion
superfluid at mean field level and can be evaluated as a matrix in
flavor space,
\begin{equation}
{\cal S}_\pi(p)= \left(\begin{array}{cc} {\cal S}_{uu}(p)&{\cal S}_{ud}(p)\\
{\cal S}_{du}(p)&{\cal S}_{dd}(p)\end{array}\right)
\end{equation}
with the elements
\begin{eqnarray}
&& {\cal S}_{uu} = {\left(i\nu_n+E_{\bf
p}^-\right)\Lambda_+\gamma_0\over
(i\nu_n)^2-(E_\pi^-)^2}+ {\left(i\nu_n-E_{\bf p}^+\right)\Lambda_-\gamma_0\over (i\nu_n)^2-(E_\pi^+)^2},\nonumber\\
&& {\cal S}_{dd} = {\left(i\nu_n-E_{\bf
p}^-\right)\Lambda_-\gamma_0\over
(i\nu_n)^2-(E_\pi^-)^2}+ {\left(i\nu_n+E_{\bf p}^+\right)\Lambda_+\gamma_0\over (i\nu_n)^2-(E_\pi^+)^2},\nonumber\\
&& {\cal S}_{ud} = {-i\Delta_\pi\Lambda_+\gamma_5\over
(i\nu_n)^2-(E_\pi^-)^2}+ {-i\Delta_\pi\Lambda_-\gamma_5\over (i\nu_n)^2-(E_\pi^+)^2},\nonumber\\
&& {\cal S}_{du} = {-i\Delta_\pi\Lambda_-\gamma_5\over
(i\nu_n)^2-(E_\pi^-)^2}+ {-i\Delta_\pi\Lambda_+\gamma_5\over
(i\nu_n)^2-(E_\pi^+)^2},
\end{eqnarray}
where $E_\pi^\pm = \sqrt{(E_{\bf p}^\pm)^2+\Delta_\pi^2}$ are
quark energies.

The momentum distributions of quarks and anti-quarks can be
calculated from the positive and negative energy components of the
diagonal propagators ${\cal S}_{uu}$ and ${\cal S}_{dd}$,
\begin{eqnarray}
n_u({\bf p})&=&n_{\bar d}({\bf
p})=\frac{1}{2}\left(1-\frac{E_{\bf p}^-}{E_\pi^-}\right),\nonumber\\
n_d({\bf p})&=&n_{\bar u}({\bf p})=\frac{1}{2}\left(1-\frac{E_{\bf
p}^+}{E_\pi^+}\right).
\end{eqnarray}

The gap equations to determine the effective quark mass $m$ and
pion condensate $\Delta_\pi$ can be obtained by minimizing the
thermodynamic potential,
\begin{equation}
\label{gap} \frac{\partial\Omega}{\partial m}=0,\ \ \
\frac{\partial\Omega}{\partial\Delta_\pi}=0.
\end{equation}
With the explicit form of the thermodynamic potential
\begin{equation}
\Omega=\frac{(m-m_0)^2+\Delta_\pi^2}{4G_s}-4N_c\int {d^3 {\bf
p}\over (2\pi)^3}\left[\zeta(E_\pi^+)+\zeta(E_\pi^-)\right],
\end{equation}
we obtain the explicit gap equations at zero temperature,
\begin{eqnarray}
\label{gap0} m-m_0 &=& 4N_cG_sm\int{d^3{\bf p}\over
(2\pi)^3}{1\over E_{\bf p}} \left({E_{\bf p}^-\over
E_\pi^-}+{E_{\bf p}^+\over
E_\pi^+}\right) ,\nonumber\\
\Delta_\pi &=& 4N_c G_s\Delta_\pi\int{d^3{\bf p}\over
(2\pi)^3}\left({1\over E_\pi^-}+{1\over E_\pi^+}\right).
\end{eqnarray}
The numerical results for $m$ and $\Delta_\pi$ are the same as in
Fig.\ref{fig6}, if we replace $\mu_B$ by $\mu_I$. The results
agree well with lattice data\cite{ISO1,ISO2,ISO3} at least at
small isospin chemical potential. For $\mu_I<m_\pi$, the ground
state is the same as the vacuum state and the isospin density
keeps zero, while for $\mu_I>m_\pi$, the pion condensate and
isospin density become nonzero. In the isospin chemical potential
region above but close to the critical value $\mu_I=m_\pi$, the
effective quark mass and pion condensate can be well described by
\begin{eqnarray}
&&\frac{m(\mu_I)}{m(0)}\simeq\frac{\langle\sigma\rangle(\mu_I)}{\langle\sigma\rangle(0)}=\left(\frac{m_\pi}{\mu_I}\right)^2,
\nonumber\\
&&\frac{\Delta(\mu_{\text
I})}{\langle\sigma\rangle(0)}=\sqrt{1-\left(\frac{m_\pi}{\mu_I}\right)^4}.
\end{eqnarray}

The explanation of the BEC-BCS crossover at zero temperature is
similar to the diquark case in Section \ref{s2}. For
$0<\mu_I/2<m(\mu_I)$, the minimum of the dispersion $E_\pi^-$ is
at $|{\bf p}|=0$ where the energy gap is
$\sqrt{\mu_N^2+\Delta_\pi^2}$. For large enough $\mu_I$, the
minimum of the dispersion is shifted to $|{\bf p}|\simeq\mu_I/2$
where the energy gap is $\Delta$. The momentum distribution for
$u$ and anti-$d$ quarks is the same as in Fig.\ref{fig7}, which
shows a BEC-BCS crossover at finite isospin chemical potential.
Similar to the diquark case, the chemical potential $\mu_I^0$ for
the crossover is
\begin{equation}
\mu_I^0=\left[2m(0)m_\pi^2\right]^{1/3},
\end{equation}
which is about $230 - 270$ MeV when the parameter values change
reasonably.

The second order effective action which is quadratic in the meson
fields and controls the meson behavior can be evaluated as
\begin{eqnarray}
S_{\text{eff}}^{(2)}[\sigma,\pi]&=&\int_0^\beta d\tau\int
d^3{\bf x}\frac{\sigma^2+\pi^2}{4G_s}\nonumber\\
&&+\frac{1}{2}\text{Tr}\left\{{\cal S}_\pi\Sigma[\sigma,\pi]{\cal
S}_\pi\Sigma[\sigma,\pi]\right\}.
\end{eqnarray}
In momentum space, it is related to the meson polarizations,
\begin{equation}
S_{\text{eff}}^{(2)}[\sigma,\pi]=\frac{1}{2}\sum_k\left[\frac{\delta_{ij}}{2G_s}-\Pi_{ij}(k)\right]\phi_i(-k)\phi_i(k)
\end{equation}
with $i,j=\sigma,\pi_+,\pi_-,\pi_0$, where the polarization
functions $\Pi_{ij}$ are defined as
\begin{equation}
\Pi_{ij}(k)=iN_c\int\frac{d^4p}{(2\pi)^4}\text{Tr}\left[\Gamma_a{\cal
S}_\pi(p+k)\Gamma_b{\cal S}_\pi(p)\right].
\end{equation}

The masses of those new eigen modes in the superfluid phase are
determined by the pole equation
\begin{equation}
\det\left[\frac{\delta_{ij}}{2G_s}-\Pi_{ij}(k_0=M,{\bf
k}=0)\right]=0.
\end{equation}
It can be shown that the neutral pion does not mix with the other
mesons and is still an eigen mode of the system, but sigma and
charged pions are strongly mixed, and it is the mixing that
results in a gapless Goldstone boson. From the proportional
relations $\Pi_{\sigma \pi_+}, \Pi_{\sigma \pi_-} \sim
m\Delta_\pi$ and $\Pi_{\pi_+\pi_-}\sim \Delta_\pi^2$\cite{NJL7},
the mixing between sigma and charged pions is very strong in the
BEC region where $m$ and $\Delta_\pi$ are both large and coexist,
but can be neglected at large isospin chemical potential.
\begin{figure}[!htb]
\begin{center}
\includegraphics[width=6cm]{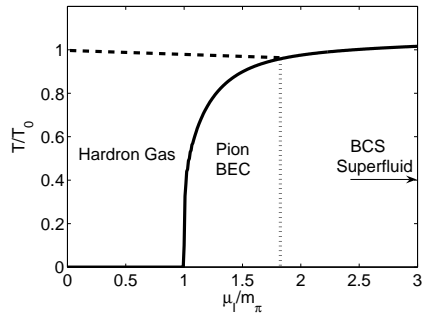}
\caption{The proposed phase diagram of pion condensation in the
$T-\mu_I$ plane. \label{fig15}}
\end{center}
\end{figure}

In summary, the phase diagram of QCD at finite isospin density is
shown in Fig.\ref{fig15}. At low temperature and low isospin
chemical potential, the matter is in normal hadron gas with chiral
symmetry breaking. With increasing temperature, there should be a
phase transition from hadron gas to quark gas, indicated by the
thick dashed line. When the isospin chemical potential becomes
larger than the pion mass in the vacuum, the pion BEC appears and
keeps until another critical isospin chemical potential $\mu_I^0$,
which is indicated by the vertical dashed line. At high enough
isospin chemical potential, the matter will enter the BCS
superfluid where the quark-antiquark Cooper pairs are condensed.
Between the BEC and BCS, there should exist a large crossover
region. Since QCD at finite isospin chemical potential can be
successfully simulated on lattice, such a BEC-BCS crossover can be
confirm by measuring the quark energy gap and comparing it with
the pion condensate, and the pseudogap phase can be confirmed by
investigating the quark spectral function. When finite baryon
density, charge neutrality and weak equilibrium are taken into
account, they may have significant effect on the BEC-BCS crossover
of pion condensation\cite{pionB1,pionB2,pionB3}.

\section {Conclusions}
\label{s5}
We have studied the BEC-BCS crossover in QCD at finite baryon or
isospin density in the NJL model at quark level. We investigated
the BEC-BCS crossover in two aspects: (1) Above the critical
temperature of the superfluid, diquarks or mesons are stable bound
states at low chemical potential but become unstable resonances at
high chemical potential; (2) At zero temperature, the effective
non-relativistic chemical potential, dispersions of fermion
excitations and the fermion momentum distribution behavior
significantly differently in the low and high chemical potential
regions. The diquark BEC-BCS crossover in two color QCD at finite
baryon density and the pion BEC-BCS crossover in real QCD at
finite isospin density can be identified, since the quark
confinement is not important in two color case. We expect that
such a crossover can be confirmed in the lattice simulations,
since the results from the NJL model agree quite well with the
lattice data obtained so far. However, for real QCD at finite
baryon density, whether there exists a diquark BEC-BCS crossover
is still an open question, since the confinement in this case is
quite important.

An interesting and important phenomenon we found in this paper is
that the BEC-BCS crossover we discussed is not induced by simply
increasing the coupling constant of the attractive interaction but
by changing the corresponding charge number. During the change of
the number density, the chiral symmetry restoration plays an
important role in the study of BEC-BCS crossover.

When an chemical potential or fermion density mismatch between the
two pairing species is turned on, the BEC-BCS crossover will be
dramatically changed. The system may go from some non-uniform
phases such as LOFF in the BCS region to some uniform gapless
phase in the BEC region\cite{NR1,NR2,NR3,sheehy,hu,he3,pieri}.

{\bf Acknowledgement:} We thank Dr. Meng Jin for helpful
discussions. The work was supported by the grants NSFC10425810,
10435080, 10575058 and SRFDP20040003103.

\begin{widetext}
\appendix
\section {Polarization Functions $\Pi_1,\Pi_2,\Pi_3,\Pi_4$}
\label{app1}
The meson and diquark polarization functions $\Pi_1,\Pi_2,\Pi_3$
and $\Pi_4$ above the critical temperature for diquark or pion
condensation can be evaluated as explicit functions of temperature
$T$ and corresponding chemical potential $\mu$,
\begin{eqnarray}
&&\Pi_1(k;\mu)=-\int\frac{d^3{\bf
p}}{(2\pi)^3}\Bigg[\left(\frac{f_{\bf p}^-+f_{\bf p}^+-f_{\bf
q}^--f_{\bf q}^+}
{k_0-E_{\bf q}+E_{\bf p}}-\frac{f_{\bf p}^-+f_{\bf p}^+-f_{\bf q}^--f_{\bf q}^+}{k_0+E_{\bf q}-E_{\bf p}}\right){\cal T}_-^-\nonumber\\
&&\ \ \ \ \ \ \ \ \ \ \ \ \ \ \ \ \ \ \ \ \ \ \ \ \ \ \ \ \
+\left(\frac{2-f_{\bf p}^--f_{\bf p}^+-f_{\bf q}^--f_{\bf
q}^+}{k_0-E_{\bf q}-E_{\bf p}}-\frac{2-f_{\bf p}^--f_{\bf
p}^+-f_{\bf q}^--f_{\bf q}^+}{k_0+E_{\bf q}+E_{\bf p}}\right){\cal
T}_+^-\Bigg]\nonumber\\
&&\Pi_2(k;\mu)=-\int\frac{d^3{\bf p}}{(2\pi)^3}\Bigg[\left(\frac{f_{\bf p}^-+f_{\bf p}^+-f_{\bf q}^--f_{\bf q}^+}
{k_0-E_{\bf q}+E_{\bf p}}-\frac{f_{\bf p}^-+f_{\bf p}^+-f_{\bf q}^--f_{\bf q}^+}{k_0+E_{\bf q}-E_{\bf p}}\right){\cal T}_-^+\nonumber\\
&&\ \ \ \ \ \ \ \ \ \ \ \ \ \ \ \ \ \ \ \ \ \ \ \ \ \ \ \ \
+\left(\frac{2-f_{\bf p}^--f_{\bf p}^+-f_{\bf q}^--f_{\bf
q}^+}{k_0-E_{\bf q}-E_{\bf p}}-\frac{2-f_{\bf p}^--f_{\bf
p}^+-f_{\bf q}^--f_{\bf q}^+}{k_0+E_{\bf q}+E_{\bf p}}\right){\cal
T}_+^+\Bigg]\nonumber\\
&&\Pi_3(k;\mu)=-2\int\frac{d^3{\bf p}}{(2\pi)^3}\Bigg[\left(\frac{f_{\bf p}^+-f_{\bf q}^-}
{k_0+\mu-E_{\bf q}+E_{\bf p}}-\frac{f_{\bf p}^--f_{\bf q}^+}{k_0+\mu+E_{\bf q}-E_{\bf p}}\right){\cal T}_-^+\nonumber\\
&&\ \ \ \ \ \ \ \ \ \ \ \ \ \ \ \ \ \ \ \ \ \ \ \ \ \ \ \ \
+\left(\frac{1-f_{\bf p}^--f_{\bf q}^-}{k_0+\mu-E_{\bf q}-E_{\bf
p}}-\frac{1-f_{\bf p}^+-f_{\bf q}^+}{k_0+\mu+E_{\bf q}+E_{\bf
p}}\right){\cal
T}_+^+\Bigg],\nonumber\\
&&\Pi_4(k;\mu)=-2\int\frac{d^3{\bf p}}{(2\pi)^3}\Bigg[\left(\frac{f_{\bf p}^--f_{\bf q}^+}
{k_0-\mu-E_{\bf q}+E_{\bf p}}-\frac{f_{\bf p}^+-f_{\bf q}^-}{k_0-\mu+E_{\bf q}-E_{\bf p}}\right){\cal T}_-^+\nonumber\\
&&\ \ \ \ \ \ \ \ \ \ \ \ \ \ \ \ \ \ \ \ \ \ \ \ \ \ \ \ \
+\left(\frac{1-f_{\bf p}^+-f_{\bf q}^+}{k_0-\mu-E_{\bf q}-E_{\bf
p}}-\frac{1-f_{\bf p}^--f_{\bf q}^-}{k_0-\mu+E_{\bf q}+E_{\bf
p}}\right){\cal T}_+^+\Bigg],
\end{eqnarray}
where we have defined ${\bf q}={\bf p}+{\bf k}$, $E_{\bf
p}=\sqrt{{\bf p}^2+m^2}$, $f_{\bf p}^\pm=f(E_{\bf p}\pm\mu/2)$
with $f(x)=\left(1+e^{x/T}\right)^{-1}$ being the Fermi-Dirac
distribution function and ${\cal T}_\pm^\mp=1\pm\left({\bf
p}\cdot{\bf q}\mp m^2\right)/\left(E_{\bf p}E_{\bf q}\right)$.

\section{Meissner Masses Squared of Gluons}
\label{app2}
In this Appendix we list the explicit form of the Meissner masses
squared $m_4^2$ for the 4-7th gluons and $m_8^2$ for the 8th gluon
in two-flavor color superconductivity phase which are quoted from
\cite{meissner1,meissner2},
\begin{eqnarray}
m^2_{8}&=&\frac{2g^2}{9}\int\frac{d^3{\bf p}}{(2\pi)^3}
\left[2\left(3-\frac{{\bf p}^2}{E_{\bf p}^2}\right)A({\bf p})+\frac{{\bf p}^2}{E_{\bf p}^2}B({\bf p})\right],\\
m_4^2&=&\frac{2g^2}{3}\int\frac{d^3{\bf p}}{(2\pi)^3}
\left[\left(3-\frac{{\bf p}^2}{E_{\bf p}^2}\right)C({\bf
p})+\frac{{\bf p}^2}{E_{\bf p}^2}D({\bf p})\right]
\end{eqnarray}
with the QCD gauge coupling constant $g$ and the functions
\begin{eqnarray}
A({\bf p})&=&\frac{(E_\Delta^-)^2-E_{\bf p}^-E_{\bf
p}^++\Delta^2}{(E_\Delta^-)^2-(E_\Delta^+)^2}\frac{
\Theta(-E_1)+\Theta(-E_2)-1}{E_\Delta^-}-\frac{(E_\Delta^+)^2-E_{\bf
p}^-E_{\bf p}^++\Delta^2}{(E_\Delta^-)^2-(E_\Delta^+)^2}\frac{
\Theta(-E_3)+\Theta(-E_4)-1}{E_\Delta^+}+\frac{1}{E_{\bf p}},\nonumber\\
B({\bf p})&=&-\delta(E_1)-\delta(E_2)
-\delta(E_3)-\delta(E_4), \nonumber\\
C({\bf
p})&=&u_-^2\left(\frac{\Theta(-E_5)-\Theta(E_2)}{E_5+E_2}+\frac{\Theta(-E_7)-\Theta(E_1)}{E_7+E_1}\right)
+v_-^2\left(\frac{\Theta(-E_5)-\Theta(-E_1)}{E_5-E_1}+\frac{\Theta(-E_7)-\Theta(-E_2)}{E_7-E_2}\right)\nonumber\\
&+&u_+^2\left(\frac{\Theta(-E_6)-\Theta(E_3)}{E_6+E_3}+\frac{\Theta(-E_8)-\Theta(E_4)}{E_8+E_4}\right)
+v_+^2\left(\frac{\Theta(-E_6)-\Theta(-E_4)}{E_6-E_4}+\frac{\Theta(-E_8)-\Theta(-E_3)}{E_8-E_3}\right)+\frac{2}{E_{\bf
p}}, \nonumber\\
D({\bf
p})&=&u_-^2\left(\frac{\Theta(-E_6)-\Theta(-E_2)}{E_6-E_2}+\frac{\Theta(-E_8)-\Theta(-E_1)}{E_8-E_1}\right)
+v_-^2\left(\frac{\Theta(-E_6)-\Theta(E_1)}{E_6+E_1}+\frac{\Theta(-E_8)-\Theta(E_2)}{E_8+E_2}\right)\nonumber\\
&+&u_+^2\left(\frac{\Theta(-E_5)-\Theta(-E_3)}{E_5-E_3}+\frac{\Theta(-E_7)-\Theta(-E_4)}{E_7-E_4}\right)
+v_+^2\left(\frac{\Theta(-E_5)-\Theta(E_4)}{E_5+E_4}+\frac{\Theta(-E_7)-\Theta(E_3)}{E_7+E_3}\right),
\end{eqnarray}
where the quark energies are defined as
\begin{eqnarray}
&& E_1=E_\Delta^- -\delta\mu,\ \ \ E_2=E_\Delta^- +\delta\mu,\ \ \
E_3=E_\Delta^+ -\delta\mu,\ \ \
E_4=E_\Delta^+ +\delta\mu,\nonumber\\
&& E_5= E_b^+-\delta\mu,\ \ \ E_6= E_b^-+\delta\mu, \ \ \ E_7=
E_b^++\delta\mu,\ \ \ E_8=E_b^--\delta\mu
\end{eqnarray}
with the chemical potential difference $\delta\mu$ between $u$ and
$d$ quarks and $E_\Delta^\pm$ and $E_b^\pm$ being listed in
Section \ref{s3}, the coherent coefficients $u_\pm^2$ and
$v_\pm^2$ are defined as $u_\pm^2=\left(1+E_{\bf
p}^\pm/E_\Delta^\pm\right)/2$ and $v_\pm^2=\left(1-E_{\bf
p}^\pm/E_\Delta^\pm\right)/2$. Note that we have added the terms
$1/E_{\bf p}$ to $A$ and $2/E_{\bf p}$ to $C$ to cancel the vacuum
contribution\cite{kitazawa2}. In this way the Meissner masses
squared are guaranteed to be zero in the normal phase with
$\Delta=0$.
\end{widetext}

\end{document}